\documentclass[apj,numberedappendix]{emulateapj}

\pdfoutput=1

\usepackage{hyperref}

\usepackage{natbib}
\bibpunct{(}{)}{;}{a}{}{,}
\usepackage{color}
\definecolor{darkgreen}{RGB}{0,142,128}

\usepackage{amsmath}
\usepackage{amsthm}
\usepackage{amsfonts}
\usepackage{url}
\usepackage{subfigure}
\usepackage{multirow}

\newcommand{\B}{\mathbf{B}}

\newcommand{\J}{\mathbf{J}}

\newcommand{\U}{\mathbf{U}}

\newcommand{\dr}{\partial_r}
\newcommand{\drr}{\partial^2_{rr}}
\newcommand{\dth}{\partial_\theta}
\newcommand{\dphi}{\partial_\varphi}

\newcommand{\bnab}{\boldsymbol{\nabla}}
\newcommand{\er}{\mathbf{e}_r}
\newcommand{\etheta}{\mathbf{e}_\theta}
\newcommand{\ephi}{\mathbf{e}_\varphi}
\newcommand{\rot}{\bnab\times}
\newcommand{\Div}{\bnab\cdot}

\newcommand{\grad}{\bnab}
\newcommand{\gradperp}{\bnab_\perp}

\newcommand{\Rlm}[1]{\mathbf{R}^{m_{#1}}_{l_{#1}}}
\newcommand{\Slm}[1]{\mathbf{S}^{m_{#1}}_{l_{#1}}}
\newcommand{\Tlm}[1]{\mathbf{T}^{m_{#1}}_{l_{#1}}}
\newcommand{\sumlm}[1]{\sum_{l_{#1}=1}^{\infty}\sum_{m_{#1} = - l_{#1}}^{l_{#1}}}
\newcommand{\sumlmn}[1]{\sum_{l_{#1}=1}^{\infty}\sum_{m_{#1} = - l_{#1}}^{l_{#1}}\sum_{\nu_{#1}=-1}^1}

\newcommand{\mysum}[3]{\sum_{\substack{l_{#1},l_{#2} = 1 \\ l_{#3} \ge |l_{#1}-l_{#2}| \\
      l_{#3}\le l_{#1} + l_{#2}}  }^{\infty}
      \sum_{\substack{m_{#1} = -l_{#1} \\m_{#2} = -l_{#2} \\ m_{#1}+m_{#2}=m_{#3}}
}^{l_{#1},l_{#2}}\sum_{\nu_{#1},\nu_{#2}} }
\newcommand{\Wlm}[1]{W^{l_{#1}}_{m_{#1}}}
\newcommand{\Zlm}[1]{Z^{l_{#1}}_{m_{#1}}}
\newcommand{\Clm}[1]{C^{l_{#1}}_{m_{#1}}}
\newcommand{\Alm}[1]{A^{l_{#1}}_{m_{#1}}}

\newcommand{\Xlmn}[1]{X_{#1;l_{#1},l_{#1}+\nu_{#1}}^{\hspace{0.1cm}m_{#1}}}
\newcommand{\Ylm}[1]{Y_{l_{#1}}^{m_{#1}}}
\newcommand{\Ylmn}[1]{\mathbf{Y}_{l_{#1},l_{#1}+\nu_{#1}}^{\hspace{0.1cm}m_{#1}}}
\newcommand{\dint}[2]{\hspace{0.2cm}\mbox{d}^{#2}{#1}}
\newcommand{\MAlm}[1]{\mathcal{A}^{l_{#1}}_{m_{#1}}}
\newcommand{\MBlm}[1]{\mathcal{B}^{l_{#1}}_{m_{#1}}}
\newcommand{\MClm}[1]{\mathcal{C}^{l_{#1}}_{m_{#1}}}

\newcommand{\Klm}[3]{#1_{l_{#2},l_{#2}#3}^{\hspace{0.1cm}m_{#2}}}

\newcommand{\modif}[1]{#1}
\newcommand{\modiff}[1]{#1}
\newcommand{\modifff}[1]{#1}
\newcommand{\modiffff}[1]{#1}
\newcommand{\referee}[1]{#1}
\newcommand{\refereee}[1]{#1}
\begin{document}

\title{Magnetic energy cascade in spherical geometry: \\
  I. The stellar convective dynamo case.}

\author{A. Strugarek, A.S. Brun, S. Mathis}
\affil{Laboratoire AIM Paris-Saclay, CEA/Irfu Universit\'e Paris-Diderot CNRS/INSU, F-91191 Gif-sur-Yvette.}
\email{antoine.strugarek@cea.fr}

\author{Y. Sarazin}
\affil{CEA, IRFM, F-13108 Saint-Paul-lez-Durance, France.}
\shorttitle{Spherical spectral transfers: magnetic energy}
\shortauthors{Strugarek, et al.}
\begin{abstract}
\modiff{We present a method to characterize the spectral transfers of magnetic energy between scales in
  simulations of stellar convective dynamos. The full triadic transfer functions are
computed thanks to analytical coupling relations of spherical
harmonics based on the Clebsch-Gordan coefficients. The method is
applied to mean field $\alpha\Omega$ dynamo
models as benchmark tests. From the physical standpoint, the decomposition of the dynamo field into primary and
secondary dynamo families proves very instructive in the $\alpha\Omega$ case. The
same method is then applied to a fully turbulent dynamo in a solar
convection zone, modeled with the 3D MHD ASH code. \modifff{The initial
  growth of the magnetic energy spectrum is shown to be
  non-local. It mainly reproduces the kinetic
  energy spectrum of convection at intermediate scales. During the saturation phase,} two kinds of direct magnetic energy cascades
are observed in regions encompassing the smallest scales involved in the
simulation. The first cascade is obtained through the
  shearing of magnetic field by the large scale differential rotation
  that effectively cascades magnetic energy. The second is a
  generalized cascade that involves a range of local magnetic
  and velocity scales.} \referee{Non-local transfers appear to be
  significant, such that the net transfers cannot be reduced to the
  dynamics of a small set of modes. The saturation of the large scale axisymmetric
  dipole and quadrupole are detailed. In particular, the dipole is saturated by
  a non-local interaction involving the most energetic scale of the
  magnetic energy spectrum, which points out the importance of the
  magnetic Prandtl number for large-scale dynamos.}
\end{abstract}


\maketitle

\section{Introduction}
\label{sec:introduction}

Magnetic fields are observed in astrophysical bodies in a
broad range of scales, from the object scale to the smallest dissipative
scales \citep{Donati:2009if}. The origin of such fields is, in most cases, due to a
hydromagnetic dynamo process. Recent developments in dynamo theory led to a distinction between
large-scale and small-scale dynamos \citep{Cattaneo:2001hg}. The large-scale dynamos produce
magnetic fields at larger scale than the largest velocity scale (or,
the largest driving scale),
while small-scale dynamos generate magnetic fields at all scales
smaller or equal to the driving scales \citep{Tobias:2011da}. Large
scale dynamos also sometimes refer to dynamos that develop a large-scale
magnetic field in super-equipartition with large scale kinetic
energy \citep{Olson:1999ev}. In that case, small-scale dynamos refer to those that develop a spectrum
peaked at small scales. Both
dynamos are generally acting together,
like in the Sun, where we observe both large-scale, intense, global
magnetic fields \citep{Schrijver:2003vu,DeRosa:2011dq,DeRosa:2012ve}, and
small-scale magnetic fields \citep{Hagenaar:2003ew,Centeno:2007cc}. In the case where multiple scales coexist also
in the velocity field, special care is needed to isolate 'small' and
'large' scales tendencies \citep{Tobias:2008fc}.\\

 Recent developments
in numerical simulations in $3$D spherical geometry allow us to model fully
non-linear dynamos in stars, involving a broad range of scales
\citep{Brun:2004ji,Browning:2008dn,Brown:2010cn,Racine:2011gh,Kapyla:2012dg}.
\refereee{The flow scales of a stellar convection zone extend from the large-scale differential
  rotation down to the smallest convective scales.}
In order to properly characterize such
dynamos, one may use specific methods to tackle the multi-scale
aspect of the problem. The principal tools that have been used in the
literature are spectral decomposition \citep{Frick:1998ke,Dar:2001gf}, and
wavelet analysis \citep{Farge:1992jr}. In this paper, we choose to use
spherical harmonics decomposition (which is adapted to the
spherical geometry of stars, \citet{Bullard:1954fo}) to develop a spectral analysis
of energy transfers in the frame of dynamo theory. 
\\

\modif{
Mainly used
to study turbulence \citep{Frisch:1995ue,Debliquy:2005jr,Lesieur:2008vl,Alexakis:2005bl},
spectral analysis is also a useful tool to characterize magnetohydrodynamic (MHD)
processes like dynamos \citep{Biskamp:1993uf,Blackman:2002cr,Mininni:2005kx,Livermore:2010dp} \referee{or
the magneto-rotational instability (MRI, see \citet{Lesur:2011jh})}.
Understanding spectral energy transfers between scales in such processes
may reinforce our ability to characterize non-linear MHD phenomena. The
shell-to-shell or mode-to-mode methods have been recently and
extensively used in the context of MHD turbulence. Indeed, the classical
Kolmogorov approach to turbulence must be adapted to the MHD
case, since the magnetic field induces an anisotropy that has to be taken
into account \citep{Iroshnikov:1964vb,Kraichnan:1965gi,Biskamp:1993uf,Goldreich:1995hq}. 
Depending on the dimensionality of the problem, spectrum slopes are
often understood to result from local (direct or indirect) transfers of
energy, referred as \textit{cascades} \citep{Biskamp:1993uf,Maron:2004bn}. However,
it was found that non-local interactions in MHD turbulence may also contribute importantly to the built-up of the
spectrum \citep{Schilling:2002dt,Aluie:2010kv}. The directions and localizations of energy transfers are then less obvious to
identify, and studies dedicated to transfer processes in spectral space
are essential to properly understand spectrum slopes in MHD \citep{Politano:1998dw,Boldyrev:2009ie,Pouquet:2011ep}.
}\\

In the past, spectral analyses have mainly been used with
Fourier spectral decomposition, generally in cartesian
coordinates and periodic parallelepipedic boxes. The Fourier
decomposition is indeed a natural way to
understand spectra, since the Fourier wave numbers represent the
inverse of a spatial scale. \refereee{More recently,
  \citet{Hughes:2012wk} used Fourier spectral analysis to study the
  influence of large-scale sheared flows on local convective dynamos. They
  show that the dynamo process depends on a broad range of scales
  in this case. We also study dynamo process in the present paper,
  though in spherical geometry 
  with a
  self-consistently generated large scale sheared flow (the
  differential rotation).}
In the case of stars or planets, the spherical geometry of
the object makes the spherical harmonics basis much more adapted to
the spectral analysis \referee{\citep{Bullard:1954fo}}. For example, \citet{Ivers:2008kp}
wrote the decomposition of the MHD equations onto spherical harmonics in the
framework of geodynamics. They were able to analytically express the non linear terms by calculating
the coupling between the spherical harmonics with the
Clebsch-Gordan coefficients \citep[see also][in a stellar
context]{Mathis:2005kz}. \refereee{The spherical harmonics
decomposition was also used by \citet{Livermore:2010dp} to develop a
spectral analysis which is similar to the one
we present in this paper. They used it to identify the spectral
interactions leading to a different saturation level of large-scale magnetic field
in kinematic and non-linear forced dynamos. In the latter case, they
observe a significant reorganization of the magnetic field such that a
strong large-scale magnetic field can emerge.}
In addition to a particular geometry, the
choice of a certain basis for the spectral analysis may be motivated
by the presence of anisotropy \modiff{(\textit{e.g.}, between the vertical and
horizontal directions)}, which is poorly described by the
classical Fourier decomposition \citep[\textit{e.g.}, see][in the case
of turbulent convection]{Rincon:2006jm}. 
\\

The spectral interactions in MHD involve triads coupling, meaning
that two modes interact to impact a third one through a triangulation
rule. Depending on the ideal MHD invariant considered,
these kinds of interaction involve couplings between the
velocity and the magnetic field, impacting the magnetic or the
velocity field. Shell-to-shell methods
generally only consider dual interactions, raising an ambiguity on the
\modiff{medium (third component)} of the triadic interaction
\citep{Verma:2005gs}. \modifff{In order to cope with this ambiguity, other
studies \citep[\textit{e.g.},][]{Schilling:2002dt} made use of the eddy-damped
quasi-neutral Markovian (EDQNM) two point statistical closure
\citep{Frisch:1975ea,Pouquet:1976eg} to get an analytical
expression of the triadic interactions. We point out that such
methods are relevant for, \textit{e.g.}, developing subgrid-scale
models for large eddy simulations (LES).
\citet{Tobias:2008kj} demonstrated that such \textit{truncated} methods often
badly describe coherent turbulent structures in flows, which are thought to be responsible for
the generation of large-scale fields in dynamos. This limitation is relevant for reduced
spectral models, which aim to reproduce the full turbulent behavior 
with a reduced number of modes. However, here we directly calculate
the full triadic shell-to-shell interactions of all the scales included in
our simulations, (\textit{i.e.}, we do not use any specific closure in
spectral space to compute the full triadic interactions).}\\

\refereee{
We here applied our spherical harmonic based method to a numerical
simulation of a stellar convection by considering the three spectral
components of the triadic interactions. 
 The originality of the
method we develop in the present work resides in the facts that \textit{(i)} we decompose \textit{explicitly} the spectral
interactions for both the magnetic and velocity fields,
\textit{(ii)} calculate explicitly all the coupling coefficients
between those fields and
\textit{(iii)} we use it to study dynamo action in a solar-like turbulent convection
zone that possesses self-consistent large-scale flows (differential
rotation, meridional circulation, ...) as well as a broad range of turbulent scales.}

\modif{
In section \ref{sec:transf-functs-spher} we present the set of MHD
equations we will use, derive from them the spectral evolution
equation for the magnetic energy in
the spherical harmonics formalism and analytically validate our
method. A toy model of an axisymmetric $\alpha-\Omega$
dynamo is analyzed with our spectral method in
Sect. \ref{sec:mean-field-kinematic}. In section \ref{sec:numer-exper}, we apply our method to study
non-linear dynamo action in a numerical simulation of a solar convective zone. Finally, conclusions and
perspectives are given in section \ref{sec:concl-persp}.
}

\section{Magnetic energy evolution equation}
\label{sec:transf-functs-spher}

\subsection{Main equations in physical space}
\label{sec:main-equat-phys}

We use the well-tested Anelastic Spherical Harmonics (ASH) code which
models turbulent stellar convection zones \citep{Clune:1999vd,Jones:2011in}. It solves the following three dimensional MHD
set of equations (see \citet{Brun:2004ji}) in the anelastic
approximation, in a reference frame rotating at the angular velocity
$\mathbf{\Omega}_0=\Omega_0\mathbf{e}_z$ (where $\mathbf{e}_z$ is the
cartesian vertical axis):
\begin{align}
 \label{eq:mass_consrv}
 \boldsymbol{\nabla}\cdot\left(\bar{\rho}\mathbf{U}\right) &= 0, \\
   \label{eq:Maxw1}
   \boldsymbol{\nabla}\cdot\mathbf{B} &= 0, \\
   \bar{\rho}\left[\partial_t \mathbf{U} + \left(\mathbf{U}\cdot \mbox{\boldmath $\nabla$}
     \right)\mathbf{U} + 2\mbox{\boldmath $\Omega_0$}\times\mathbf{U} \right] &=-\mbox{\boldmath $\nabla$} P +
   \rho \mathbf{g}  \nonumber \\
   +\frac{1}{4\pi}\left(\mbox{\boldmath $\nabla$}\times\mathbf{B}\right)\times\mathbf{B} &-
   \mbox{\boldmath $\nabla$}\cdot\mathbf{D} - \left[\mbox{\boldmath $\nabla$}\bar{P}-\bar{\rho}\mathbf{g} \right],
          \label{eq:mom_conserv} \\
          \bar{\rho}\bar{T}\left[\partial_t S +
         \mathbf{U}\cdot\mbox{\boldmath
           $\nabla$}\left(\bar{S}+S\right) \right] &=
              \boldsymbol{\nabla}\cdot\left[
                \kappa_r\bar{\rho}c_p\mbox{\boldmath $\nabla$}\left(\bar{T}+T\right)\right. \nonumber
                  \\ 
                  +  \kappa_{0}\bar{\rho}\bar{T}\mbox{\boldmath $\nabla$}\bar{S} &+
                  \left.  \kappa\bar{\rho}\bar{T}\mbox{\boldmath $\nabla$} S \right] +
                  \frac{4\pi\eta}{c^2}\mathbf{J}^2 \nonumber \\ 
                  + 2\bar{\rho}\nu [ e_{ij}e_{ij}
                  &-\frac{1}{3}\left(\boldsymbol{\nabla}\cdot\mathbf{U}\right)^2
                  ] \, ,
   \label{eq:entropy_conserv} \\
  \partial_t \mathbf{B} =
   \mbox{\boldmath
     $\nabla$}\times\left(\mathbf{U}\times\mathbf{B}\right) & -
   \mbox{\boldmath $\nabla$} \times\left(\eta\mbox{\boldmath $\nabla$}\times\mathbf{B}\right),
   \label{eq:induction_eq}    
\end{align}
where the \modiff{spherically symmetric background} thermodynamical state is denoted by bars (fluctuations
with respect to the background state are denoted without bars), $\mathbf{v}$ is the local
velocity, $\kappa_r$ is the radiative diffusivity, and $\kappa$, $\nu$ and $\eta$
 are respectively the effective thermal diffusivity, the eddy
 viscosity and the magnetic diffusivity. The thermal diffusion coefficient $\kappa_0$
applies at the top of the modeled convective zone (where convective
motions vanish), to ensure the heat transport through the
\modiff{upper} surface. 
$\mathbf{J}=(c/4\pi)\mbox{\boldmath $\nabla$}\times\mathbf{B}$
is the current density, and the viscous stress tensor $\mathbf{D}$
is defined by
\begin{equation}
  \label{eq:visc_stress_tensor}
  D_{ij} = -2\bar{\rho}\nu\left[e_{ij}-\frac{1}{3}\left(\mbox{\boldmath $\nabla$}\cdot\mathbf{U}\right)\delta_{ij}\right] \, ,
\end{equation}
where $e_{ij}$ is the strain rate tensor, and $\delta_{ij}$ is the
Kronecker symbol. The system is closed by using the linearized ideal gas law:
\begin{equation}
  \label{eq:gas_law}
  \frac{\rho}{\bar{\rho}} = \frac{P}{\bar{P}} - \frac{T}{\bar{T}} =
  \frac{P}{\gamma \bar{P}} - \frac{S}{c_p}
\end{equation}
with $c_p$ the specific heat at constant pressure and $\gamma$ the
adiabatic exponent. 
The vectorial fields are decomposed in poloidal and toroidal
components:
\begin{eqnarray}
  \label{eq:B1}
  \B(r,\theta,\varphi) &=& \rot\rot\left[C(r,\theta,\varphi)\er\right]
  \nonumber \\
  &+& \rot\left[A(r,\theta,\varphi)\er\right], \\
  \label{eq:U1}
  \bar{\rho}(r)\U(r,\theta,\varphi) &=&
  \rot\rot\left[W(r,\theta,\varphi)\er\right] \nonumber \\
  &+& \rot\left[Z(r,\theta,\varphi)\er\right]\, ,
\end{eqnarray}
where $(\mathbf{e}_r,\mathbf{e}_\theta,\mathbf{e}_\varphi)$ are the
unit vectors in spherical coordinates. \modifff{All the quantities are
  time-dependent}. This decomposition ensures numerically that both
the magnetic field and the mass flux remain
divergenceless up to the machine precision. \\
A potential match of the magnetic field ($\bnab\times\mathbf{B} = 0$)
is applied both at the bottom and top \modiff{radial} boundaries. For the convective dynamo case
(Sect. \ref{sec:numer-exper}), the boundary conditions for the
velocity are impenetrable and stress-free. A latitudinal entropy
gradient is imposed at the bottom (as in \citet{Miesch:2006iz}), and we
fix a constant entropy gradient at the top of the domain.

\subsection{Magnetic energy transfer functions}
\label{sec:magn-energy-transf}

\subsubsection{The formalism}
\label{sec:formalism}

In this section, we present a method to obtain a spectral (in the
sense of the spherical harmonics) evolution equation for the
magnetic energy, starting from the induction equation \eqref{eq:induction_eq}.
In order to deal with vectorial fields and spherical harmonics (see
Eq. \eqref{eq:norm} in Appendix), it is practical to define the vectorial spherical
harmonics basis \citep{Rieutord:1987go,Mathis:2005kz}:
\begin{equation}
  \label{eq:RST}
  \left\{
  \begin{array}{lcl}
    \Rlm{}(\theta,\varphi) &=& \Ylm{}(\theta,\varphi) \er \\
    \Slm{}(\theta,\varphi) &=& \gradperp \Ylm{} = \dth\Ylm{}\etheta + \frac{1}{\sin{\theta}}\dphi\Ylm{}\ephi\\
    \Tlm{}(\theta,\varphi) &=& \gradperp\times\Rlm{} = \frac{1}{\sin{\theta}}\dphi\Ylm{}\etheta  -\dth\Ylm{}\ephi
  \end{array}
  \right. .
\end{equation}
It is an orthogonal basis for the scalar product $\int_S \cdot
\dint{\Omega}{}$, where $S$ is a spherical surface and $\mbox{d}\Omega =
\sin\theta\mbox{d}\theta\mbox{d}\varphi$
the associated infinitesimal solid angle. \refereee{The mode numbers $m$
  and $l$ are the azimuthal wave number and the spherical harmonic
  degree (which characterize to their latitudinal variations).
}
The general properties of this basis maybe found in appendix
\ref{sec:leftrlm-slm-tlmright}. 
The two main vectorial fields that appear in the induction equation
\eqref{eq:induction_eq} are the magnetic and the velocity fields. We
want to project those fields on the vectorial basis \eqref{eq:RST},
using the decompositions \eqref{eq:B1}-\eqref{eq:U1}. Fortunately, the
curl of a vector is a linear operation that can be expressed very
easily in the vectorial spherical harmonics basis (see
equation \eqref{eq:curl}). We obtain:
\begin{align}
  \B(r,\theta,\varphi) &= \sumlm{} \left\{
    \frac{l(l+1)}{r^2}\Clm{}(r)\Rlm{} \right.  \nonumber \\ &+
      \frac{1}{r}\dr\Clm{}(r)\Slm{} + \left. \frac{\Alm{}(r)}{r}\Tlm{}      \right\},
  \label{eq:B2}
  \\
    \bar{\rho}(r)\U(r,\theta,\varphi) &=  \sumlm{} \left\{
      \frac{l(l+1)}{r^2}\Wlm{}(r)\Rlm{} \right. \nonumber \\ &+
      \frac{1}{r}\dr\Wlm{}(r)\Slm{} + \left. \frac{\Zlm{}(r)}{r}\Tlm{}
    \right\},  \label{eq:U2} 
\end{align}
where
we have projected the toroidal and poloidal components of the fields
($A$, $C$, $Z$ and $W$) on the
scalar spherical harmonics basis. \modif{We see that in equation
  \eqref{eq:B2}, the poloidal $C$ and toroidal $A$ components of $\B$
  are respectively projected on $(\mathbf{R},\mathbf{S})$ and
  $\mathbf{T}$. Consequently, in the remainder of this paper the projection of any vectorial field on
  $(\mathbf{R},\mathbf{S})$ will be referred as \textit{poloidal}, and
  the projection on $\mathbf{T}$ as \textit{toroidal}.} 

\subsubsection{Shell to shell analysis}
\label{sec:shell-shell-analysis}

To study the transfers of
energy between scales on a spherical surface, we
distinguish the different scales of the axisymmetric ($m=0$) and
non-axisymmetric physical fields
by defining shells $L^0$ and $L^\star$ as follows:
\begin{eqnarray}
  \label{eq:spect_shell_axi}
  \mathbf{X}_{L}^0 &=& \mathcal{A}^l_0\mathbf{R}^0_l +
  \mathcal{B}^l_0\mathbf{S}^0_l + \mathcal{C}^l_0\mathbf{T}^0_l, \\
  \label{eq:spect_shells}
  \mathbf{X}_{L}^\star &=& \sum_{\substack{-l\le m \le l \\ m \ne 0}} \left\{ \mathcal{A}^l_m\Rlm{} + \mathcal{B}^l_m\Slm{}
  + \mathcal{C}^l_m\Tlm{} \right\}.
\end{eqnarray}
\modiff{This distinction is natural when studying the generation
  of large scale axisymmetric field. 
  Another choice of shells based on
  dynamo families will also be
used in this paper (see Appendix \ref{sec:prim-second-famil} and the end
of Sect. \ref{sec:spectr-magn-energy}
).}
Note that the defined shells are orthogonal, \textit{i.e.} that any
scalar product of strictly different shells is zero. In order to
simplify the notations, the shell
$L$ may represent either axisymmetric or non-axisymmetric shells. Exponents $^0$ and
$^\star$ denote axisymmetric and non-axisymmetric components in
the remainder of this paper. \\
The shells of magnetic energy in spectral space are then defined by
\begin{equation}
  \label{eq:ME_sp}
  E_L^{\rm mag} = \frac{1}{2}\int_S \mathbf{B}_L \cdot \mathbf{B}_L \dint{\Omega}{}.
\end{equation}
This spectrum may be defined at any radial location of the spherical
domain (\textit{e.g.}, in a stellar interior). Our spectral analysis
intends to characterize horizontal
scales of velocity and magnetic fields, and \refereee{can be easily
  applied at any and as many as necessary}
depths, depending on the radial locations one wants to
focus on. \refereee{In addition, the different terms of the evolution
  equation of the magnetic energy (see next section) explicitly depend upon the radial
  gradients of the different quantities. The horizontal
  couplings produced by the vertical interactions are thus taken into
  account by our description. Finally, it is worth noting that the
  spectra may be different at various depths in a convective dynamo
  model. In this work, we will only study pure convection zone
  dynamos, we will consequently focus on the spectral interaction in
  the middle of the convection zone.}

\subsubsection{Spectral magnetic energy equation}
\label{sec:spectr-magn-energy}

In order to obtain the spectral
magnetic energy evolution equation, we multiply equation (\ref{eq:induction_eq}) by $\mathbf{B}_L$ and integrate
it over the spherical surface so that
\begin{equation}
  \label{eq:ME_terms}
 \partial_t E^{\rm mag}_L = \mathcal{D}_L +
 \sum_{L_1,L_2} \left\{ \mathcal{P}_L\left(L_1,L_2\right) +
   \mathcal{F}_L\left(L_1,L_2\right)  \right\}\, ,
\end{equation}
where $\mathcal{D}$ regroups the diffusion terms, $\mathcal{P}$
represents the volumetric
production of magnetic energy, and $\mathcal{F}$ is the
\modiff{divergence of the} flux of magnetic
energy through a spherical surface. \modiff{Note that the production
  $\mathcal{P}$ term has to be understood as a general production
  term, that can either be positive (real production) or negative (destruction).} The sum over $L_1, L_2$ involves
the triangular selection rule $\left\{\left| l_1-l_2\right| \leq l \leq
l_1+l_2,\, m_1+m_2=m\right\}$ that comes naturally from the spherical
harmonics coupling (see 
Appendix \ref{sec:mathbfym_l-l+nu-basi}).
The expressions of the three
contributions $\mathcal{D}, \, \mathcal{P}$, and $\mathcal{F}$ are given by:
\begin{align}
  \label{eq:Diff_term}
  &\mathcal{D}_L(r) = \int_S \left\{\eta\B_L\cdot\nabla^2\B_L
  \right. \nonumber \\
  & \left. + \dr\eta\er\cdot\left(\B_L\times\rot\B_L \right)\right\} \dint{\Omega}{}, \\
   &\mathcal{P}_L\left(r,L_1,L_2\right) = \int_S
   \left(\U_{L_1}\times\B_{L_2}\right)\cdot\rot\B_L \dint{\Omega}{} ,
  \label{eq:Prod_term}
   \\
   &\mathcal{F}_L\left(r,L_1,L_2\right) =
     \int_S \Div\left[ \left(\mathbf{U}_{L_1}\times\mathbf{B}_{L_2}\right)\times\B_L \right] \dint{\Omega}{}.
  \label{eq:Flux_term}
\end{align}
We have split the diffusive term by considering 
a
diffusivity $\eta$ 
that
only depends on $r$.  
The production and flux of magnetic energy are discretized
over scales so that we compute which scale of the velocity field
($L_1$) is interacting with which scale of the magnetic field ($L_2$)
towards a studied scale $L$. 

\referee{Although the expressions \eqref{eq:Diff_term}-\eqref{eq:Flux_term} are
formally written, they include the evaluation of vectorial products 
decomposed on the vectorial spherical harmonics basis. 
This
operation is not easily calculated in the
$(\mathbf{R},\mathbf{S},\mathbf{T})$ basis 
thus 
we use
an alternative basis \citep{Varshalovich:1988ul} to compute it. 
For sake of simplicity, these details are given
in appendix \ref{sec:couplings-mag}.}

\referee{
Alternatively to considering axisymmetric and non-axisymmetric spectra, it
is instructive to decompose the flow and field into the so-called
\textit{primary} (dipolar, antisymmetric) and
\textit{secondary} (quadrupolar, symmetric) families
\citep{McFadden:1991bw,Roberts:1972wu}. These families were proven
very insightful to characterize geophysical and astrophysical dynamos
\citep{Gubbins:1993iia,DeRosa:2011dq,DeRosa:2012ve}. Further,
the primary/secondary distinction greatly simplifies the transfer maps of
$\mathcal{P}$ and $\mathcal{F}$. Indeed, the coupling between 
fields of the same family always gives \textit{secondary} fields, while the
coupling between fields of different families always gives a
\textit{primary} field (see Appendix
\ref{sec:prim-second-famil}). \modiff{Both of these approaches
  (\textit{i.e.}, axisymmetric/non-axisymmetric and primary/secondary
  distinctions) will be used hereafter.}
}

\subsection{Validation and illustration of the method}
\label{sec:valid-meth-simple}

\modif{In this section, we illustrate the coupling calculations for two
  simple fields. The reader only interested in physical discussions
  may skip this part and go directly to Sect. \ref{sec:mean-field-kinematic}.\\}
Since the ASH code is a spectral code, it solves the MHD equations for
the spherical harmonics coefficients of the fields. Although it does
not compute explicitly the decomposition on the vectorial spherical
harmonics basis \eqref{eq:RST}, it is straightforward to make use of this basis in
the code by using the transformation relations
\eqref{eq:B2}-\eqref{eq:U2}. We have added in the code the ability to
compute the different terms \eqref{eq:Diff_term}-\eqref{eq:Flux_term}
of the spectral magnetic energy equation \eqref{eq:ME_terms}.

\begin{figure}[htbp]
\includegraphics[width=9cm]{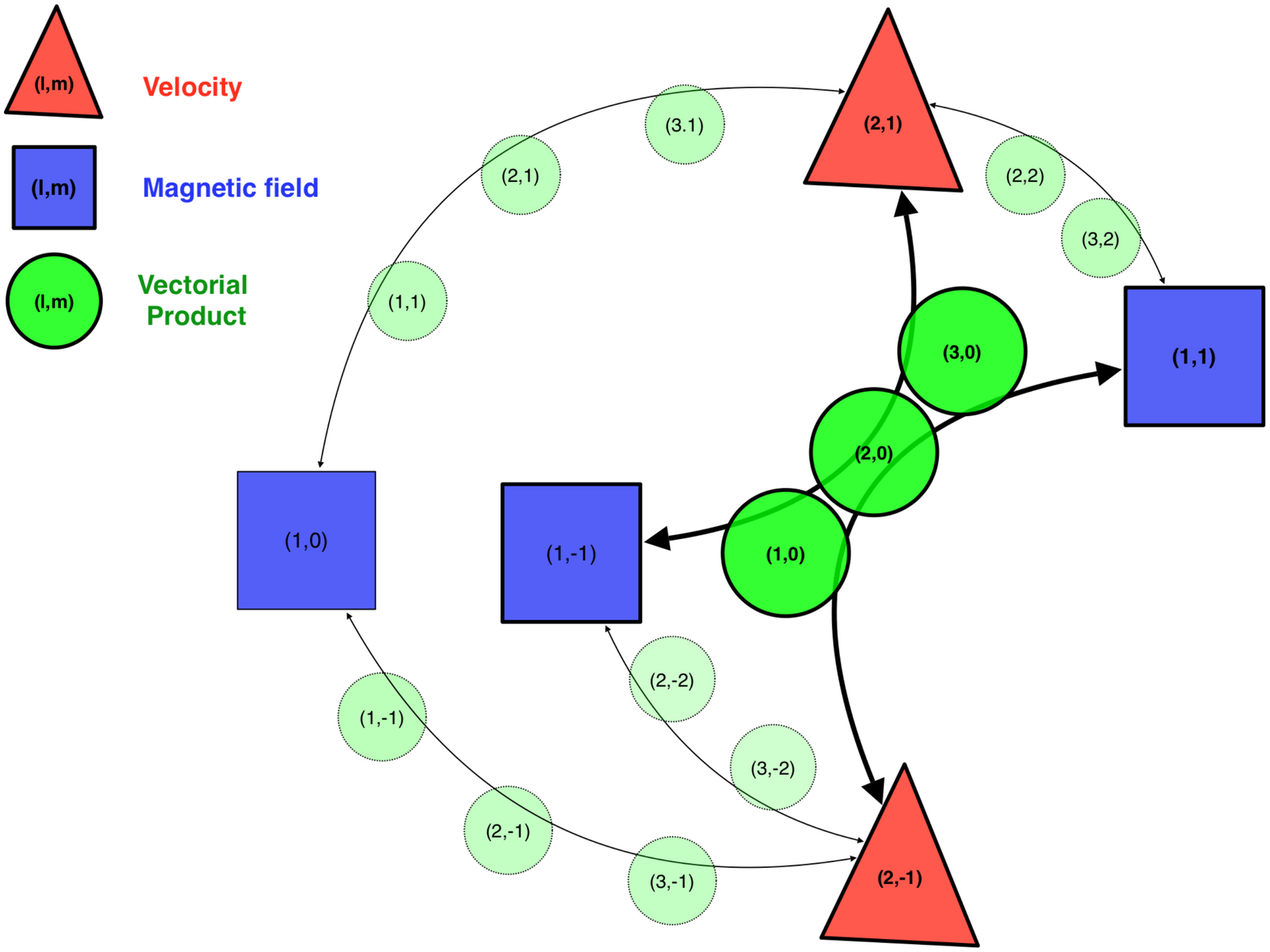}
\caption{Schematic coupling between spherical harmonics of
  $\mathbf{U}$ and $\mathbf{B}$ for the simple test case. The
  $\mathbf{B}$ modes are represented by blue squares, and the
  $\mathbf{U}$ modes by red triangles. The black arrows represent the
  coupling between the modes, the green circles on them represent
  resulting modes obtained from the coupling \textit{via}
  the triangulation rule of 
  the vectorial product $\mathbf{U}\times\mathbf{B}$. The three highlighted
  green circles in the center correspond to the modes calculated in appendix \ref{sec:numerical-validation}.}
\label{fig2}
\end{figure}

In order to \referee{illustrate and} validate both the coupling coefficients \eqref{eq:vect_prod}
for the vectorial product and the general method, we numerically computed
a simple analytical test case. We
initialize the magnetic and velocity fields in the following way:
\begin{eqnarray*}
  \mathbf{B} &=& a\left( \mathbf{R}_1^{0} +
    \frac{1}{2}\mathbf{R}_1^{1} - \frac{1}{2}\mathbf{R}_1^{-1} \right)
   + b \left( \mathbf{S}_1^{0} +
    \frac{1}{2}\mathbf{S}_1^{1} - \frac{1}{2}\mathbf{S}_1^{-1} \right) \\
  &+& c \left( \mathbf{T}_1^{0} +
    \frac{1}{2}\mathbf{T}_1^{1} - \frac{1}{2}\mathbf{T}_1^{-1}
  \right), \\
  \bar{\rho}\mathbf{U} &=& d\left(
    \mathbf{R}_2^{1} - \mathbf{R}_2^{-1} \right) + 
    e \left(
    \mathbf{S}_2^{1} - \mathbf{S}_2^{-1} \right) + f\left(
    \mathbf{T}_2^{1} - \mathbf{T}_2^{-1} \right)
\end{eqnarray*}
where $a,b,c,d,e$ and $f$ are functions of $r$ only.
This initialization allows us to test at the same time the
axisymmetric/non-axisymmetric and non-axisymmetric/non-axisymmetric
coupling schemes between the velocity and the magnetic fields. The low
order harmonics ($l\in\left\{0,1,2\right\}$) that are involved make the analytic
calculation \modifff{easy}. We display on figure \ref{fig2} the
possible couplings (\textit{via} vectorial product) between the
$\mathbf{U}$ and $\mathbf{B}$ fields we initialized. This is in fact a
schematic representation of the triangulation rule that appear in the
summation of equation \eqref{eq:vect_prod}. The analytical
calculation of the values of the three \modiff{large green} circles is given in 
appendix
\ref{sec:numerical-validation}. \referee{The resulting vectorial products
calculated by the code using the Wigner coefficients show very good
agreement with the coefficients calculated analytically (table
\ref{tab:tab1} in appendix \ref{sec:numerical-validation}).}

\modiff{
We stress here
that this test has been done for low $m$ and $l$ values. The numerical accuracy
of the algorithms calculating the Clebsch-Gordan coefficients (and thus
3j, 6j and 9j Wigner coefficients) is known to decrease with increasing $l$ and
$m$. The calculation routines we use are accurate up to values of $l$ of the
order of $500$. To do so, we used a multiple precision
package\footnote{\url{http://crd-legacy.lbl.gov/~dhbailey/mpdist/}} to
simulate large-precision numbers that are needed to compute the
ratios of factorials and binomial coefficients that are involved in the
Wigner coefficients calculations. However, the calculation time of the transfer functions $\mathcal{P}$ and $\mathcal{F}$
increases dramatically with $l$ and $m$. For practical reason, when
computing fully nonlinear dynamos (see Sect. \ref{sec:numer-exper}
below), we have chosen to limit the computation of the coupling
coefficients to $l_{\rm max}=70$, even if the effective resolution of
such simulations reaches $l_{\rm max}=340$.
From time to time, we do calculate the transfer terms
for high $l$'s to have an indication of how  energy is transfered
at the smallest scales (see
Sect. \ref{sec:numer-exper}). Nevertheless, the
magnetic-energy-carrying scales in the spectrum are
dominated by $l\le 70$ in this case. We thus capture the essential
part of the dynamics.
}

\section{Axisymmetric $\alpha\Omega$ dynamo}
\label{sec:mean-field-kinematic}

In this section, we use the spectral method we developed in
Sect. \ref{sec:transf-functs-spher} on two academic cases. First, we
explain how the classical $\Omega$ effect \citep{Moffatt:1978tc} is
represented by our formalism (Sect. \ref{sec:omega-effect-}). Then, we calculate the spectral
transfers for a mean field $\alpha\Omega$ model (Sect. \ref{sec:mean-field-axisymm}).

\subsection{Omega effect}
\label{sec:omega-effect-}

\referee{
The complexity of the two spherical harmonics bases may be confusing
when it comes to interpret simple and classical dynamo processes. We
thus give hereafter a step-by-step explanation of the $\Omega$-effect
in the two vectorial spherical harmonics bases formalism. \\
We start with a purely dipolar poloidal magnetic field that reads
(using Eq. \eqref{eq:curl})
\begin{eqnarray}
  \mathbf{B}_p(r,\theta) &=& 
  b_r(r)\mathbf{R}^0_1 + b_\theta(r)\mathbf{S}^0_1\, . 
  \label{eq:BP_Omega_effect}
\end{eqnarray}
Then, we want to calculate the effect of a differential rotation that
reads
\begin{equation}
  \Omega = A + B\cos^2\theta\, .
  \label{eq:BP_Omega_choice}
\end{equation}
Such differential rotation is usually seen as a ``$l=2$''
field. Though, it projects on a $l=3$ component when considering
the azimuthal component of the velocity $\mathbf{U}_\varphi =
r\sin\theta\Omega\, \mathbf{e}_\varphi$ (see \citet{Roberts:1972wu}),
which reads
\begin{eqnarray}
\label{eq:DR_Omega_effect_true}
  \mathbf{U}_\varphi(r,\theta) &=& U_\varphi(r)\left(A\sin\theta +
    B\sin\theta\cos^2\theta\right)\mathbf{e}_\varphi   \\ 
&\sim& U_\varphi(r)\partial_\theta
  Y^0_3\mathbf{e}_\varphi = U_\varphi(r)\mathbf{T}^0_3 \, . 
\label{eq:DR_Omega_effect}
\end{eqnarray}
In general,
Eq. \eqref{eq:DR_Omega_effect_true} should project both on $\mathbf{T}^0_1$
and $\mathbf{T}^0_3$. For the sake of simplicity, we select here a
profile of differential rotation that is purely described by a
$(l=3,m=0)$ harmonic, \modifff{which corresponds to $B=-5A$ (Eq. \eqref{eq:BP_Omega_choice}).} 
We simply apply the curl operator \eqref{eq:curl} and make use of the
coupling relations \eqref{eq:vect_prod} to obtain
the production of $\mathbf{B}$ in the induction equation,
\begin{align}
  \label{eq:Omega_effect_end}
  \bnab&\times\left(\mathbf{U}_\varphi\times\mathbf{B}_p\right) =
  f\left(b_r,b_\theta,U_\varphi\right)
\mathbf{T}^0_2\, .
\end{align}
We recovered that the action of differential rotation on a purely
axisymmetric poloidal field creates a toroidal field $B_\varphi
\propto \sin\theta\cos\theta$. With our notations, this kind of
field will be labeled as a '$l=2$' field.
}

\referee{
An additional feature of the differential rotation can also be learnt
from this little analysis. We immediately remark that for axisymmetric
fields, the first Wigner coefficient involved in the coupling between
two shells $L_1$ and $L_2$ is zero if $3+l_1+l_2$ is odd (equation
\eqref{eq:J_clebsch}), 
\textit{i.e.}, if $l_1$ and $l_2$ are of the opposite parity. The
shearing effect of differential rotation will then always couple
axisymmetric scales of the magnetic field that are of opposite parity,
which will be observed in the transfer maps in more complex cases
(\textit{e.g.}, Figs \ref{fig:evol_dom_term_AO}(b) and \ref{fig:large_scale_sat}). 
}

\referee{
This simple example
strikingly highlights how the vectorial product formula
\eqref{eq:vect_prod} couples together two simple fields. \modiff{This
description of the $\Omega$ effect will guide our analysis in
Sects. \ref{sec:mean-field-axisymm} and \ref{sec:numer-exper}.}
}

\begin{figure*}[htbp]
  \centering
    \includegraphics[width=0.6\linewidth]{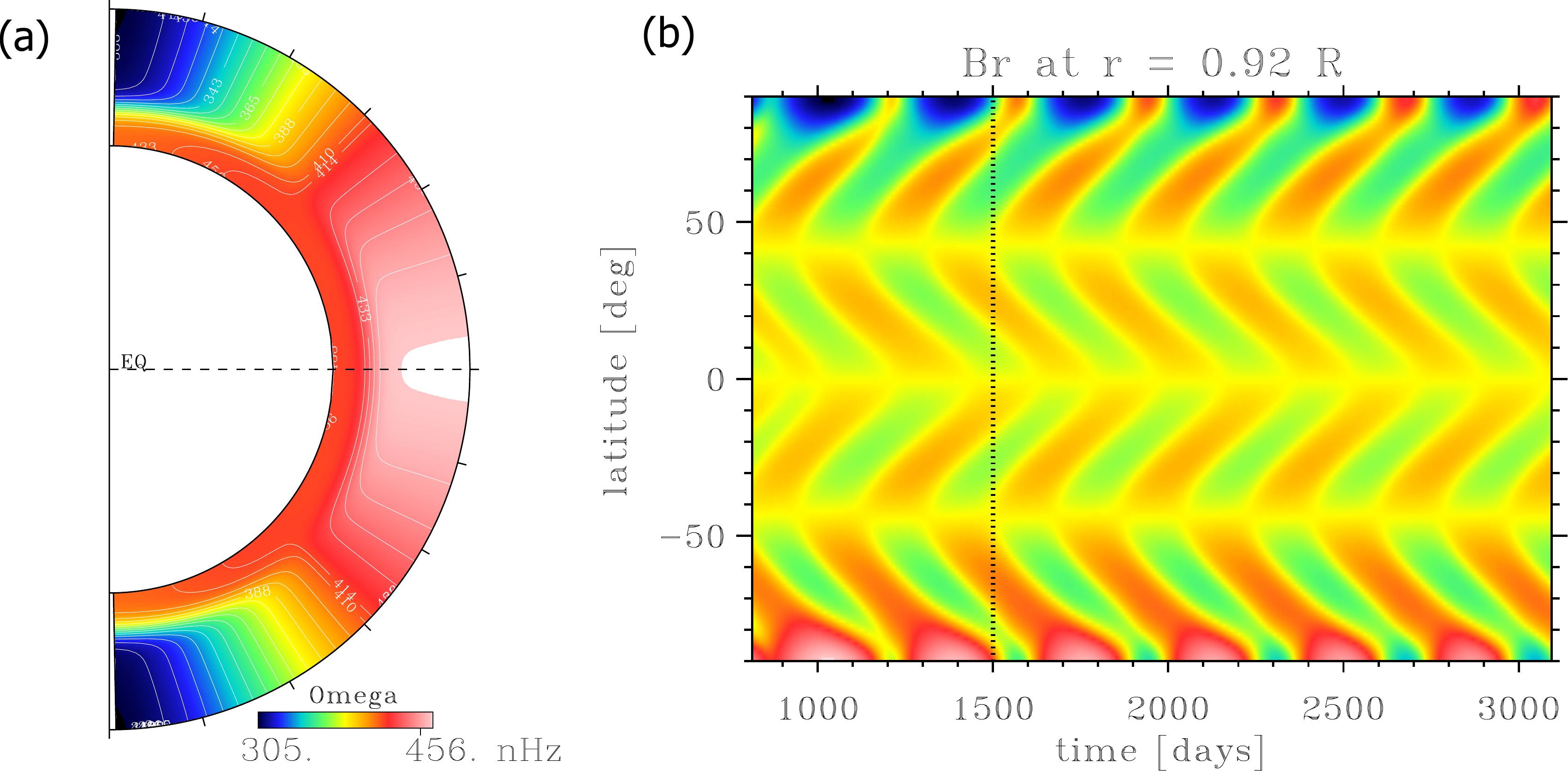}
  \caption{\refereee{\textbf{(a)} Solar like differential rotation profile computed from
    \citet{Schou:1998bi}. The tachocline is located near $r=0.7\,
    R_\odot$, and the base of the tachocline rotates at the solar
    rotation rate $\Omega_0 = 2.6\, 10^{-6}\, s^{-1}$. \textbf{(b)}
    Butterfly diagram in the axisymmetric
    $\alpha-\Omega$ dynamo (only $B_r$ as a function of time and
    latitude is shown, in the upper convection zone). The
    vertical black dotted line represent the time at which we display
    the spectral interactions in Fig. \ref{fig:evol_dom_term_AO}.}
  }
  \label{fig:DR_But_S98}
\end{figure*}

\subsection{Case of a cyclic mean field dynamo}
\label{sec:mean-field-axisymm}

We use the ASH code \citep{Clune:1999vd,Brun:2004ji} to simulate an axisymmetric mean field 
dynamo \citep{Charbonneau:2010tw,Jouve:2008fh}. \refereee{To do so, we solve
only the induction equation considering uniquely a prescribed
differential rotation profile \citep[see also section 3.8 of][for a similar use
of a 3D spherical code to model $\alpha-\Omega$
dynamos]{Jouve:2008fh}.}

Our radial domain is defined between $r_b
= 0.6\, R_\odot$ and $r_t = 0.966\, R_\odot$. We use a resolution of
$ N_r \times N_\theta \times N_\varphi = 64\times 128 \times 256$. 
\refereee{We choose a solar differential rotation profile
  $\Omega_{DR}(r,\theta)=-\partial_\theta Z_{DR} / (r^2\sin\theta)$
through the toroidal component $Z_{DR}$ of the momentum,
which is, in the frame rotating at $\Omega_0 = 2.6\, 10^{-6}\, s^{-1}$:
\begin{equation}
  \label{eq:z_dr_choice}
  Z_{DR}(r,\theta) = Z_t(r)\left( A\cos\theta +
    \frac{B}{3}\cos^3\theta + \frac{C}{5}\cos^5\theta \right)\, .
\end{equation}
From \citet{Schou:1998bi}, we take $A=257$ nHz, $B=321$ nHz and
$C=529$ nHz. The differential rotation then naturally projects on
$\U^0_1$, $\U^0_3$ and $\U^0_5$. The radial profile $Z_t(r)$ is chosen
such as to simulate a stable region at the base of the
domain and is defined by
\begin{equation}
  \label{eq:zt_definition}
  Z_t(r) = \bar{\rho}\frac{r^2}{2}\left[ 1 +
    \tanh{\left(\frac{r-4.87\,10^{10}}{2\,10^9}\right)} \right]\, .
\end{equation}
}

We initialize our magnetic field with
a \modif{seed $l=3$ poloidal (antisymmetric and axisymmetric)}
field.
Finally, we add an $\alpha$
effect to the induction equation such that
\begin{equation}
  \label{eq:Induction_plus_alpha_effect}
  \partial_t \mathbf{B} =
   \mbox{\boldmath $\nabla$}\times\left(\mathbf{U}\times\mathbf{B} +
     \alpha B_\varphi\mathbf{e}_\varphi\right) - \mbox{\boldmath
     $\nabla$}\times\left(\eta\mbox{\boldmath
       $\nabla$}\times\mathbf{B}\right)\, .
 \end{equation} 
\refereee{Since we do not take into account in this simple case the
  feedback of the Lorentz force on the flow via the Navier-Stokes
  equations, since we only solve the induction equation,  
  we need to quench the $\alpha$ effect.} 
Hence, $\alpha $ is defined by 
\begin{eqnarray}
  \label{eq:alpha_effect}
  \alpha(r,\theta) &=& \alpha_0 \, e^{\left(-\frac{r-0.75\,
      R_\odot}{0.05\,
      R_\odot}\right)^2}  \, \frac{\cos\theta}{1+(|\mathbf{B}|/B_{q})^2} \, .
\end{eqnarray}
This is the simplest $\alpha$ that is needed to trigger an
oscillating solar-like dynamo \citep{Charbonneau:2010tw}; it is
anti-symmetric with respect to the equator. The radial
profile of $\alpha$ is localized near the
base of the convection zone and the quenching value is given by
$B_q=10^3$ G.
\modiff{We have deliberately chosen an $\alpha$-effect that operates only on the poloidal component
  of the induction equation, therefore computing an $\alpha\Omega$ mean
  field dynamo \citep{Moffatt:1978tc}. This $\alpha\Omega$ dynamo exhibits the
characteristic butterfly diagram showed in
Fig. \ref{fig:DR_But_S98}(b) (at $r=0.92\, R_\odot$).} Although this $\alpha$ profile is
\textit{ad-hoc} and one among the many profiles that were tested in the literature
\citep[\textit{e.g.},
][]{Roberts:1972wu,Charbonneau:1997km,Bonanno:2002ck,Zhang:2003fz,Jouve:2008fh},
we chose
this form because it easily triggers an oscillatory dynamo and its effect
in spectral space can be easily calculated. 
It is consequently a good choice to illustrate our new spectral
method. \modif{With the parameters we chose, the cycle period is of
  the order of $400$ days (see Fig. \ref{fig:DR_But_S98}(b)).}


The extra $\alpha$ effect adds a new term in the spectral energy
equation \eqref{eq:ME_terms} that can lead to complex formula in
spectral space. We rewrite the energy equation
\begin{eqnarray}
 \partial_t E^{\rm mag}_L = \mathcal{D}_L +
 \int_S\bnab\times\left(\alpha
   B_\varphi\mathbf{e}_\varphi\right)_L\cdot\mathbf{B}_L\dint{\Omega}{} 
 \nonumber \\
 + 
 \sum_{L_1,L_2} \left\{ \mathcal{P}_L\left(L_1,L_2\right) +
   \mathcal{F}_L\left(L_1,L_2\right)  \right\}  \, .
  \label{eq:energy_spectral_plus_alpha}
\end{eqnarray}
\refereee{The interested reader may read Appendix
 \ref{sec:expr-alpha-effect} for a complete spectral description of this $\alpha$
  effect.}

\refereee{Wherever $|\mathbf{B}|$ is not too large, the quenching part of the
$\alpha$ effect is negligible. 
In that case, the
$\alpha$ effect \modiff{(which restores poloidal field from toroidal field)} simply couples
 a $L$ shell \modiff{of toroidal field} to its neighboring shells of
 \modiff{poloidal field}, namely $L-1$ and $L+1$. When $|\mathbf{B}|/B_q$
 becomes large, the $\alpha$ effect is quenched and the poloidal
 magnetic field stops being restored. When it is sufficiently low,
 $\alpha$ stops being quenched and the poloidal field grows
 again. This sets up a simple feedback mechanism and a cycle establishes.
}

\refereee{The magnetic energy spectrum is dominated by a $L=2$ component that
sets the phase of the total cycle. The various shells energy oscillate
with roughly the same period, but are generally out of phase. 
This phase shift is a natural ingredient that allow the reversal of
the overall field polarity \citep{Knobloch:1998ev,Tobias:2002il}. }

\begin{figure*}[htbp]
  \centering
   \includegraphics[width=0.8\linewidth]{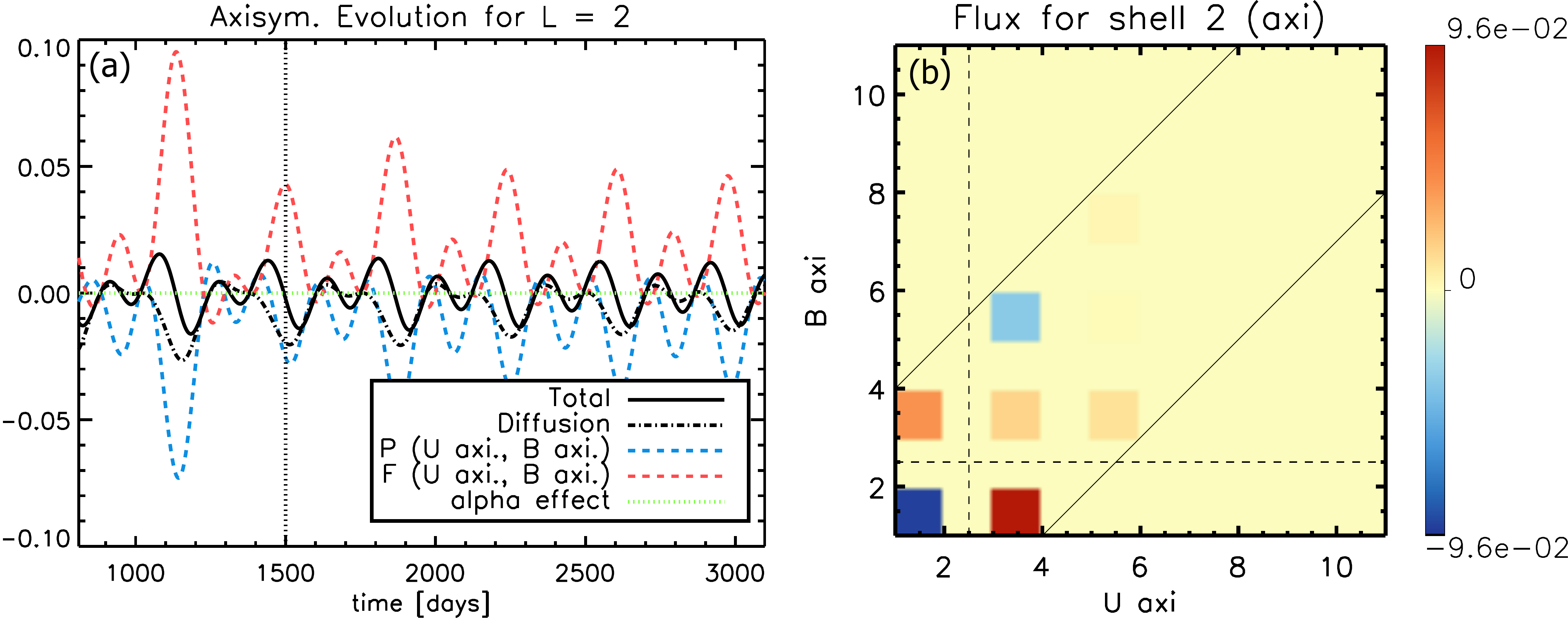}
  \caption{
    \refereee{
      \textbf{(a)} Evolution of the different terms of equation
      \eqref{eq:energy_spectral_plus_alpha} for the shell $L=2$.
      Production $\mathcal{P}$ is the dashed blue line, flux
      $\mathcal{F}$ the dashed red line, diffusion
      $\mathcal{D}_1+\mathcal{D}_2$ the dotted black line and the
      $\alpha$ effect is the dash-dot green line. The black plain line is
      the total of all the contributions. \textbf{(b)} Flux $\mathcal{F}$ contribution
    to the $L=2$ shell (see
    Eq. \eqref{eq:energy_spectral_plus_alpha}) at $t=1500$ days
    (vertical dotted line in panel (a)).
    The 2D color maps are the $\B^0-\U^0$
    transfer functions \modiff{(dark red is the maximum value,
      black the minimum)}. The oblique black lines represent the boundaries of the
    triangular selection rule.
}
}
  \label{fig:evol_dom_term_AO}
\end{figure*}


\refereee{We stress here that the magnetic field created in this experiment is of the
primary family ($(\mathbf{R}^0_{2l+1},\mathbf{S}^0_{2l+1})$ and
$\mathbf{T}^0_{2l}$, see appendix \ref{sec:prim-second-famil}). Our
initial magnetic field is a poloidal primary field
($\mathbf{R}^0_3,\mathbf{S}^0_3$). As a 
result, the toroidal field created through the $\Omega$-effect is also a
primary field ($\mathbf{T}^0_2$, see section \ref{sec:omega-effect-}). Then, our
$\alpha$-effect, that creates poloidal field from toroidal field,
transforms the primary toroidal field into a primary poloidal
field $(\mathbf{R}^0_{1,3},\mathbf{S}^0_{1,3})$,
which is of the same type
than our initial magnetic field. Hence, no secondary field can be
created in the simulation (which is confirmed by our results), and the $\alpha$-effect can only act on the primary
toroidal field to create a primary poloidal field. This
is a direct consequence of the well-known separability property of the
induction equation between the dipolar and quadrupolar families, when
symmetric flows and antisymmetric $\alpha$ effect are chosen \citep{Gubbins:1993iia}.
}


\refereee{We display in Fig. \ref{fig:evol_dom_term_AO}(a) the evolution of the
different terms of the magnetic energy equation
\eqref{eq:energy_spectral_plus_alpha} for $L=2$ at $r=0.92\, R_\odot$ during the same time period than
the butterfly diagram in Fig. \ref{fig:DR_But_S98}(b). The primary
toroidal field energy clearly evolves due to the 
production $\mathcal{P}$ (dashed blue line) and flux $\mathcal{F}$
(dashed red line) terms that account for the effect of
differential rotation on the magnetic
field (as expected, since there are no other production nor advection
terms). The two terms cancel each other out with a small time-lag,
their sum combines with the ohmic diffusion (dotted line) to produce
oscillations (solid line) of the total $L=2$ energy. Note that the
$\alpha$ effect plays no role et $r=0.92\,R_\odot$ since it is
concentrated at the base of the ``convection zone'' (equation \eqref{eq:alpha_effect}).
}

\refereee{
Our new method allows us to characterize how scales interact to
produce this behavior. We display in
Fig. \ref{fig:evol_dom_term_AO}(b) the transfer map for the flux
$\mathbf{F}$ term of equation \eqref{eq:energy_spectral_plus_alpha}
for the $L=2$ shell at its maximum. The differential rotation is
composed of the 
$\U^0_1$, $\U^0_3$ and $\U^0_5$ shells
(eq. \eqref{eq:z_dr_choice}). The transfer maps during minima (not shown
here) are qualitatively opposite, which means that all
the couplings between the shells reverse sign during the cycle. This
reversal of all shells is a simple, direct consequence of the reversal
of the whole magnetic field. At this position, the poloidal magnetic
energy (not shown here) evolves because of the ohmic diffusion of the
$\alpha$-driven poloidal field at deeper radii. Here, the poloidal
magnetic field couples with the differential rotation to transfer
energy to the toroidal $L=2$ magnetic shell. The $\U^0_3-\B^0_1$
appears to be the dominant interaction that sets the $L=2$
cycle. Interestingly, we will recover this feature in the turbulent
(convective) dynamo described in section \ref{sec:numer-exper} (see
Fig. \ref{fig:large_scale_sat}(b).)
}

\referee{This $\alpha\Omega$ dynamo provides a simple
  example of how our diagnostic may be interpreted in the context of stellar dynamo.
  \refereee{Based on how our diagnostic highlights the saturating
    properties of the solar differential rotation in an $\alpha-\Omega$ case, we now} 
  apply it to a turbulent dynamo triggered in a
  stellar convection zone 
  \refereee{that also exhibit} a solar-like differential rotation profile.}

\begin{figure*}[htbp]
  \centering
  \includegraphics[width=\linewidth]{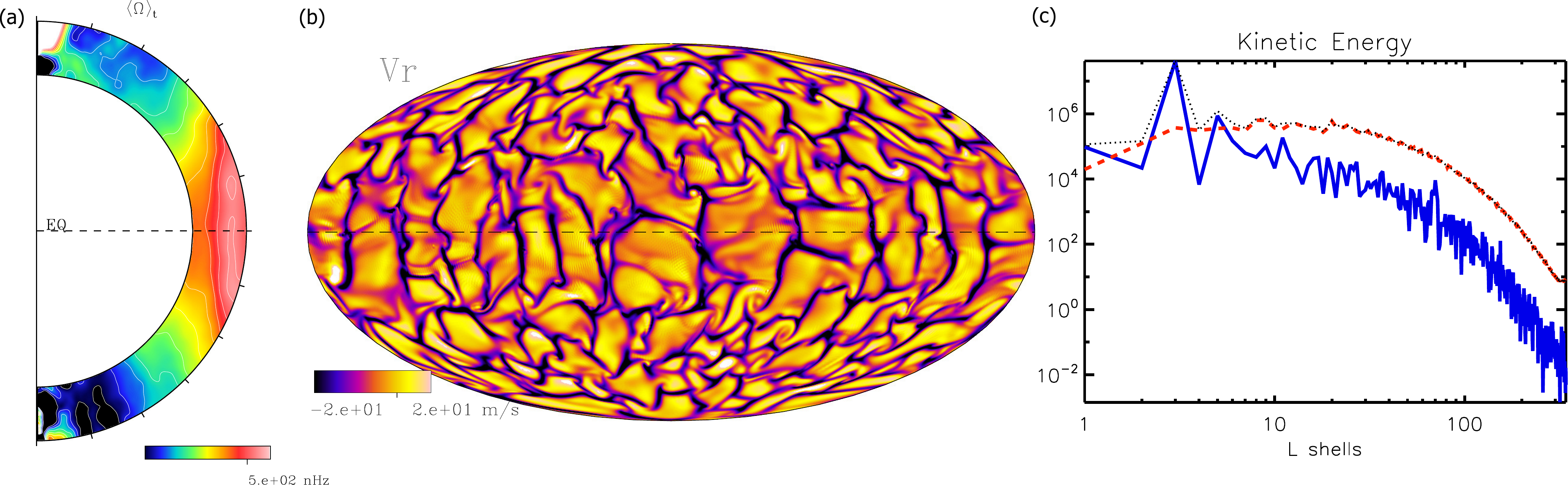}
  \caption{Progenitor hydrodynamical state. \textbf{(a)} Mean differential rotation profile averaged over
  3 months. \textbf{(b)} Time-dependent convective patterns at the top of the
  convection zone with dark tones representing downflows. \textbf{(c)} Kinetic energy spectra in the middle
  of the convection zone. The axisymmetric
    spectra are in plain blue, the non-axisymmetric spectra in dashed red
    and the total spectra in dotted black.}
  \label{fig:AB3_hydro}
\end{figure*}

\section{Nonlinear convective dynamo}
\label{sec:numer-exper}

We use the general method described in Section
\ref{sec:transf-functs-spher} and validated in Section
\ref{sec:mean-field-kinematic} to study dynamo action in a 
global (spherical) nonlinear
convection zone. \refereee{Contrary to Section
  \ref{sec:mean-field-kinematic}, we now solve the full set of MHD
  equations and do not introduce any $\alpha$ effect.} We model a turbulent solar convection zone 
\citep{Brun:2004ji,Jouve:2009jk,Pinto:2012tbs} that develops a solar-like differential rotation
profile (Fig. \ref{fig:AB3_hydro}(a)), 
with fast equator and slow poles. We
display the convective patterns we obtain in Fig. \ref{fig:AB3_hydro}(b). 
We recover the well-known 'banana'-shaped cells at the equator, and more
patchy patterns at higher latitudes. \referee{Our choice of parameters yields a
mildly turbulent state (based on the maximum amplitude of the
velocity, the Reynolds number in the middle of the convection zone is
of the order of $800$).}

\begin{figure*}
  \centering
    \includegraphics[width=\linewidth]{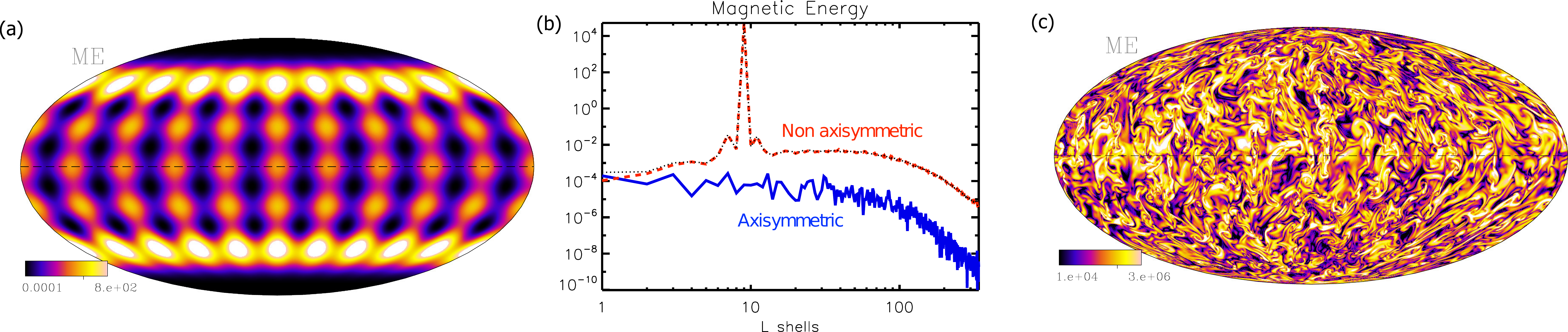}
  \caption{\textbf{(a)} Initial magnetic energy \refereee{(the seed poloidal
    streamfunction is a $l=9, m=5$ spherical harmonic)} in
    \textit{cgs}, in the middle of the convection zone. \textbf{(b)}
    Magnetic energy spectra \modiff{one time
    step after the introduction} of the peaked
    magnetic field in our turbulent convection zone. The axisymmetric
    spectra are in plain blue, the non-axisymmetric spectra in dashed red
    and the total spectra in dotted black. \modiff{\textbf{(c)}
    Saturated magnetic energy in \textit{cgs}, 600 days after the
    introduction of the seed magnetic field. \modifff{The color scale is logarithmic}.}}
  \label{fig:AB3_mag}
\end{figure*}

We display in Fig. \ref{fig:AB3_hydro}(c) 
the kinetic energy
spectra in the rotating frame at the center of the convection zone as a function of the
shell $L$. We separate the
axisymmetric component (the plain blue line) from the non-axisymmetric component
(the dashed red line), and the dotted black line is the total
spectrum. \modiff{Notice that two peaks at $L=3,5$ dominate the
kinetic energy spectrum. They represent the differential rotation
\referee{of the azimuthal
component of the toroidal velocity} (\textit{see} Sect. \ref{sec:omega-effect-}).}\\

We initialize a peaked $(l,m)=(9,5)$ non-axisymmetric magnetic field
(Fig. \ref{fig:AB3_mag}(a))  
throughout the convection
zone by setting:
\begin{equation}
  \label{eq:initB}
  \mathbf{B} = \frac{10\, B_0 R_\odot^2}{r^2}
  \left(\frac{R_b}{r}\right)^9 \mathbf{R}_9^5\, - \frac{B_0 R_\odot^2}{r^2}\left(\frac{R_b}{r}\right)^9\mathbf{S}^5_9 .
\end{equation}
We set $B_0=100$ G so the initial magnetic energy contained in the $L=9$
shell is comparable to the kinetic energy at that scale (see Fig. \ref{fig:AB3_mag}(b)). 
 
The magnetic Prandtl number throughout the convection zone
is set to $P_m = \frac{\nu}{\eta} =4$, leading to a magnetic Reynolds number of
the order of $3200$ at mid-convection zone \refereee{(based on the maximum
amplitude of the velocity)}. Such a set of
parameters triggers a dynamo instability and the growth of magnetic
energy (see Fig. \ref{fig:evol_shells_nax} in the following).

\refereee{The initialization we chose allows us to
  directly see how a significant amount of energy can be transfered to
large scales.
We also did the same numerical experiment varying the initial
conditions. By initializing roughly the same amount of energy
distributed over the whole scales, we obtained the same statistical
saturated state. Hence, this proves that in this case, the initial scale is
forgotten when the dynamo saturates.}

The complex interactions between the convective motions and the
initially peaked magnetic field lead to the construction of the 
magnetic energy spectrum. \modiff{The saturated magnetic energy after
  600 days of evolution is displayed on
  Fig. \ref{fig:AB3_mag}(c)
  in physical space, in the middle of the
  convection zone. In the remainder of this section, we
  characterize how such a state is obtained, and maintained.} We
distinguish two regimes: the development
of the spectrum shape (\refereee{the kinematic regime, }Sect. \ref{sec:creation-me-spectrum}), and its
saturation and sustainment (\refereee{the non-linear regime, }Sects. \ref{sec:saturation} and
\ref{sec:satur-mean-large}). \refereee{The results of this section will be
summarized in Fig. \ref{fig:scheme_summary}.} \modifff{We
  recall here that no $\alpha$ effect has been added to the induction
  equation \eqref{eq:induction_eq}, dynamo action is naturally
  achieved since convection is 3D and $R_m > R_m^{\rm crit}$ \citep{Brun:2004ji}.}

\subsection{Creation of magnetic energy spectrum:\\ kinematic phase}
\label{sec:creation-me-spectrum}

\begin{figure}[htbp]
  \centering
  \subfigure[]{
 \includegraphics[width=6cm,angle=90]{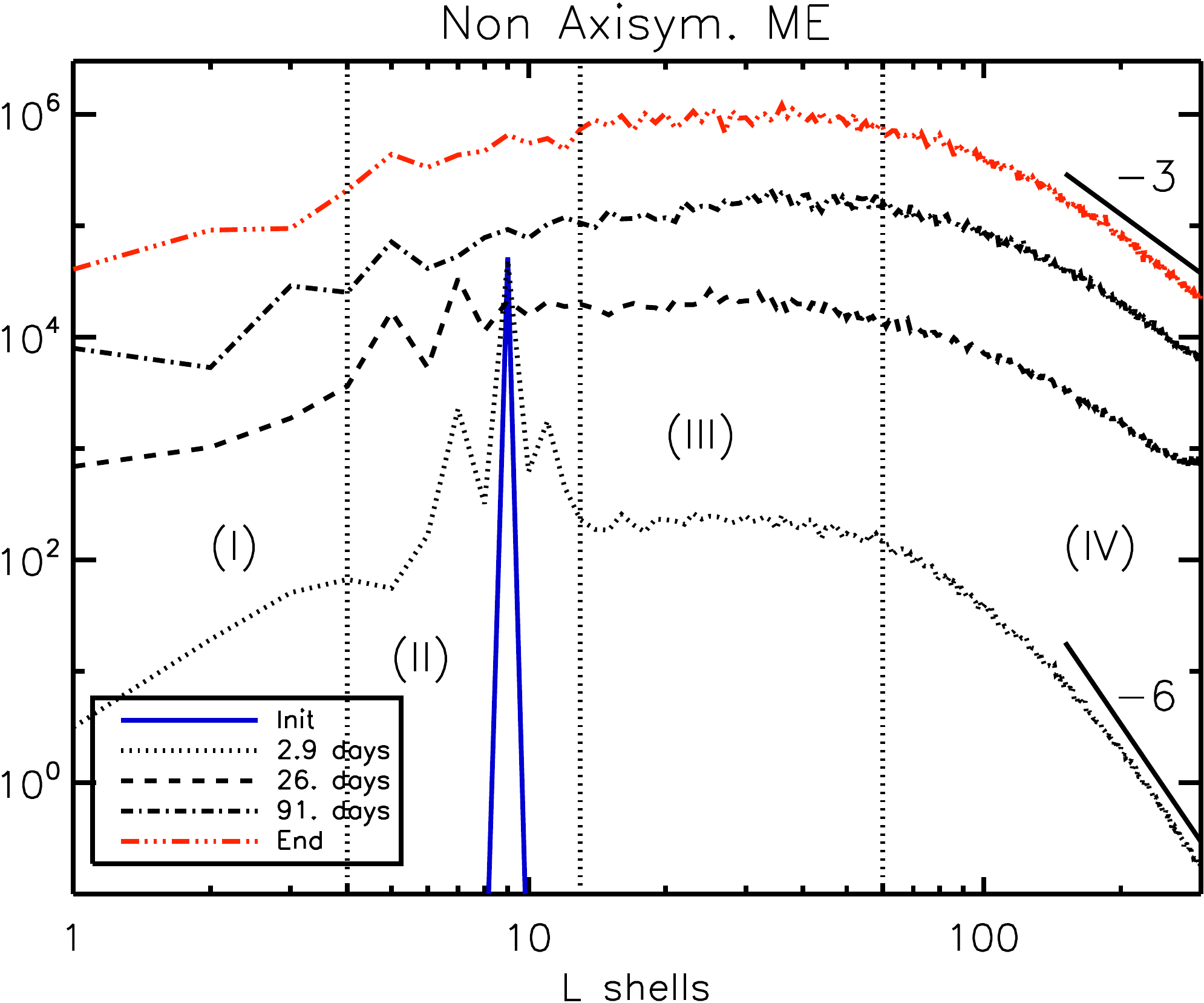}
  \label{fig:evol_ME_sp}
}
  \subfigure[]{
 \includegraphics[width=6cm,angle=90]{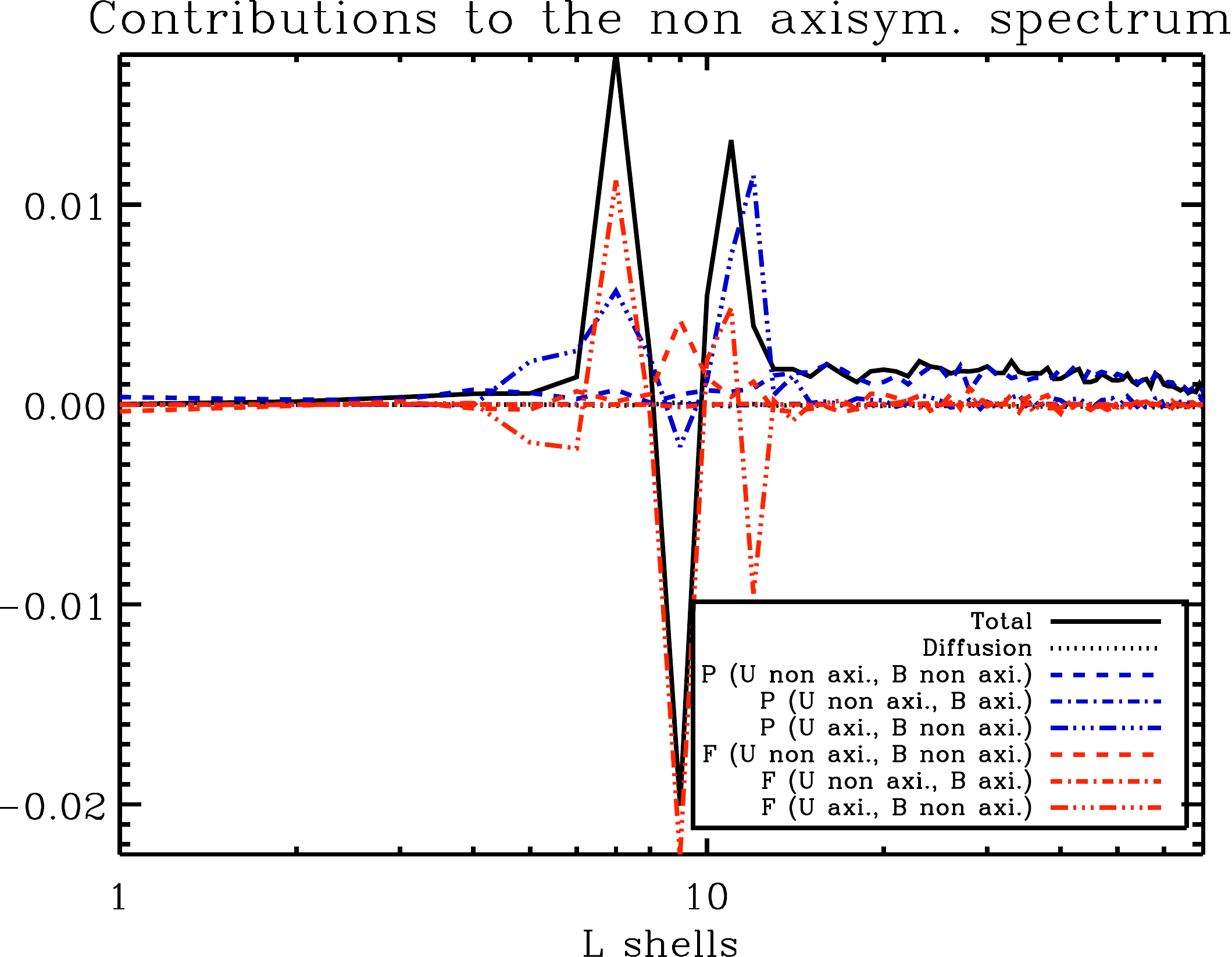}
  \label{fig:evol_ME_sum_terms}
}
  \subfigure[]{
\includegraphics[width=\linewidth]{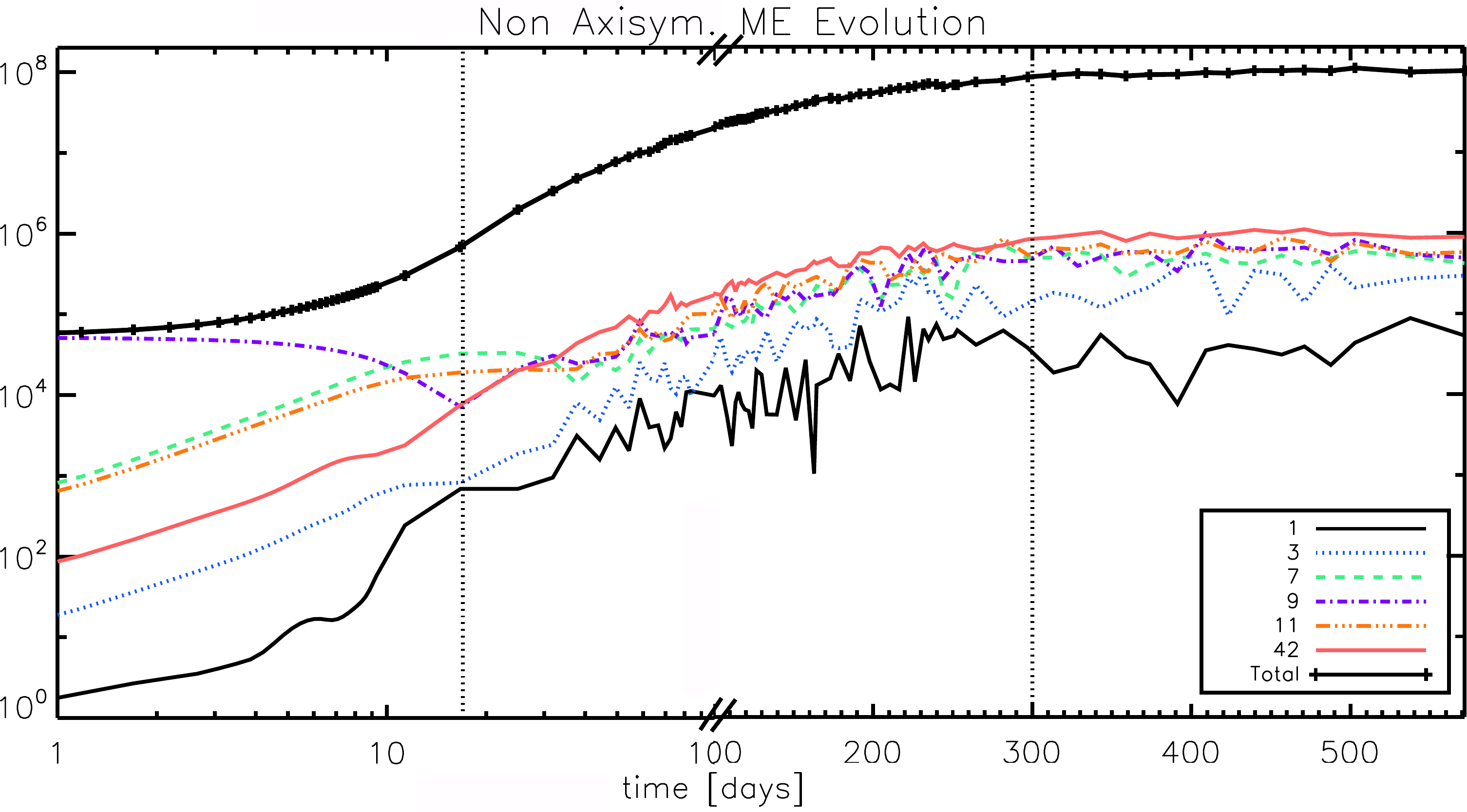}
  \label{fig:evol_shells_nax}
}
\caption{\textbf{(a)} Evolution of the non-axisymmetric magnetic
  energy spectra. The initial spectra is blue, the saturated
  spectrum red. Two slopes ($L^{-6}$ and $L^{-3}$) are given as
  references for the small scales.\textbf{(b)} Contributions to
    the non-axisymmetric magnetic energy evolution in the initial
    phase (dotted line in panel \textbf{(a)}). The total is in plain black,
  $\mathcal{D}$ in dotted black, $\mathcal{P}$ in blue and
  $\mathcal{F}$ in red.  \textbf{(c)} Evolution of
  non-axisymmetric energy of shells 1, 3, 7, 9, 11 and
  42. \referee{Notice the fast early evolution of $E_7^\star$ and
    $E_{11}^\star$ due to shearing of $\B^\star_9$ by the differential
  rotation.}  The thick plain
    line is the total non-axisymmetric energy. \referee{The abscissa is in log
    scale between for $t<100$ days, and in linear scale for $t>100$
    days. Even though the initial growth is exponential, we
    chose to represent it in log-log scale to make it appear clearly
    in the evolution plot.} }
  \label{fig:evol_sps_non_axi}
\end{figure}

We plot in Fig. \ref{fig:evol_ME_sp} the evolution of the non-axisymmetric
magnetic energy spectrum. The initial spectrum is plotted in blue, and
the saturated spectrum in red. In addition, we display in 
Fig. \ref{fig:evol_shells_nax} the
evolution of the magnetic energy for 6 different $L$ shells. The total
energy evolution is also shown (plain thick line). \referee{The initial
evolution ($t<100$ days) is shown in logarithmic scale The saturation
of magnetic energy is reached at $t\sim 300$ days.} 

\refereee{We also ran another numerical experiment where we artificially
  suppressed the Lorentz force and the ohmic heating in the momentum
  and energy equations (\textit{i.e.}, effectively running a kinematic
  dynamo). On average, the relative difference with the fully
  non-linear case starts being significantly different (departure of
  order one) roughly $10$ days after the introduction of the magnetic field
(the exact length of the kinematic phase depends on the scale
considered).
We detail hereafter how the non-axisymmetric (Sect. \ref{sec:evol-non-axisymm}) and the
axisymmetric (Sect. \ref{sec:evol-axisymm-spectr}) spectra are
created during these first days, which we will refer to as
the kinematic phase.}   

\subsubsection{Creation of the non-axisymmetric spectrum}
\label{sec:evol-non-axisymm}

\modifff{We observe at first that all the $L$ shells gain energy (Fig. \ref{fig:evol_ME_sp}), excepts the
$L=9$ shell which looses energy because it is redistributed throughout
the whole domain by the convective flows (Fig. \ref{fig:evol_shells_nax}).} \modiff{It stops decaying
 at $t\sim 17$ days.} We identify four regions in the non axisymmetric
spectrum that exhibit different behaviors. We
define the large-scale zone (I) by $1 \leq L \leq 4$, the
neighborhood zone (II) by $4 \leq L \leq 13$, and the plateau zone (III) by
$13 \leq L \leq 60$. \modif{The small-scale zone (IV) ($L \gtrsim 60$) starts at the
  highest diffusive scale, which is the highest viscous scale $l\sim
  60$ based on the first scale at which the local Reynolds number is
  lower than $1$. It also includes the magnetic dissipative scales ($L
  \gtrsim 120$)}. 
The four zones are separated by the three dotted vertical lines
in Fig. \ref{fig:evol_ME_sp}. In
order to understand how the spectrum is built, we display the contributions from the different
terms of Eqs. \eqref{eq:diff1_full_expr}-\eqref{eq:flux_full_expr} in
Fig. \ref{fig:evol_ME_sum_terms}. Those contributions are taken
shortly after the introduction of the magnetic field. They correspond
to the spectrum plotted with a dotted line in Fig.
\ref{fig:evol_ME_sp}. We recall that we fully calculate all the coupling
terms up to $L=70$.

\begin{figure*}[htbp]
  \centering
\includegraphics[width=16cm]{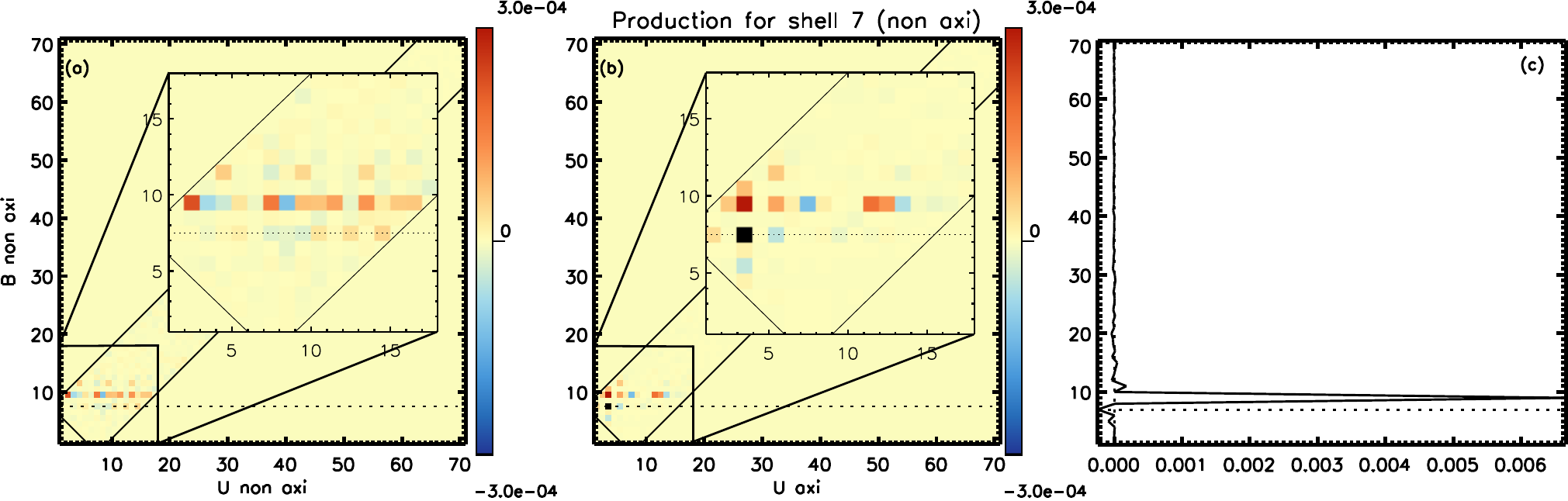}
\includegraphics[width=16cm]{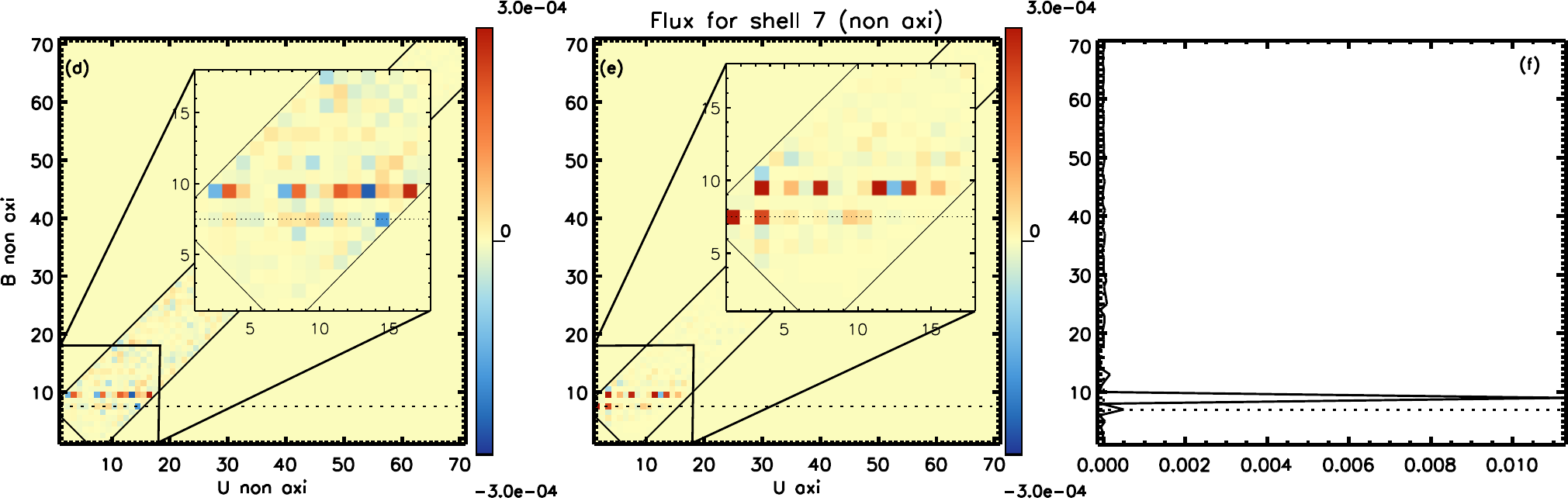}
\caption{Production $\mathcal{P}$ and flux $\mathcal{F}$ contributions
to the non-axisymmetric shell $L=7$. The 2D color maps are the $\B^\star-\U^\star$ and
$\B^\star-\U^0$ transfer functions, the 1D plot is the sum of the transfer
functions over the $\U$ shells. The horizontal dotted line labels the
$\B_7$ shell. The oblique black lines represent the boundaries of the
triangular selection rule.}
  \label{fig:prod_L7_nax_init}
\end{figure*}

\referee{
The energy transfers around the $L=9$ shell (the neighborhood zone
II) are dominated by $\U^0-\B^\star$
interactions from both the production $\mathcal{P}$ and the flux
$\mathcal{F}$ terms. \modifff{We recall here that both $\mathcal{P}$
  and $\mathcal{F}$ represent generic transfer functions, that can
  either be positive or negative.} \modifff{Dissipation is negligible
  in zone (II), even for the $L=9$ shell that
  initially contains the energy.} The $L=9$ energy then decreases through the
interaction of $\B^\star_9$ and the differential rotation $\U^0_3$
that shears the magnetic field (see Sect. \ref{sec:omega-effect-}). The energy is preferentially
redistributed to $E^\star_{11}$ and $E^\star_7$. For those two shells, the
production $\mathcal{P}$ and flux $\mathcal{F}$ terms contribute
positively to the creation of the spectrum (Fig. \ref{fig:evol_ME_sum_terms}). We display in
Fig. \ref{fig:prod_L7_nax_init} the detailed contribution of
$\mathcal{P}$ and $\mathcal{F}$ to $E^\star_7$. We only display
contributions from $\B^\star$ because the axisymmetric magnetic energy is
very small initially. The $\mathbf{U}^\star-\mathbf{B}^\star$ interactions are displayed in panel (a), and the
$\mathbf{U}^0-\mathbf{B}^\star$ interactions in panel (b). We sum over the
velocity shells to plot the production term against $\mathbf{B}^\star$ in
panel (c). We observe that the summed contribution
is dominated by $\B^\star_9-\U^0_3$ interactions, as expected.
\modiff{Also, we observe that energy is
  directly transfered from $E_9$ to $E_7^*$, such that the $L=8$
  shell is not involved in the transfer. This is true for all the
  shells in zone (II) and implies that the transfer of energy is
  non-local, even for shells close to the initial energetic shell.}
}

\modifff{Due to the triangular selection rule, the $\U^0_3-\B^\star_9$
interaction can only act in zone (II). Indeed, $L$ must be strictly greater
than $12$ in zones (III-IV) and strictly lower than $6$ in
zone (I). $\U^0_3$ and $\B^\star_9$ initially dominate respectively the kinetic
and magnetic energy spectra. Their interaction was consequently
dominant in zone (II), and we expect a different kind of spectral
transfers in the other zones. }\referee{This zone exists because of
our choice of initial condition. The very early evolution would have
been changed if we had chosen a different initial shell. Though, \refereee{as
stated before, }this
initial scale is forgotten when the saturated state is reached (Fig. \ref{fig:evol_ME_sp}).}

\begin{figure*}[htbp]
  \centering
\includegraphics[width=16cm]{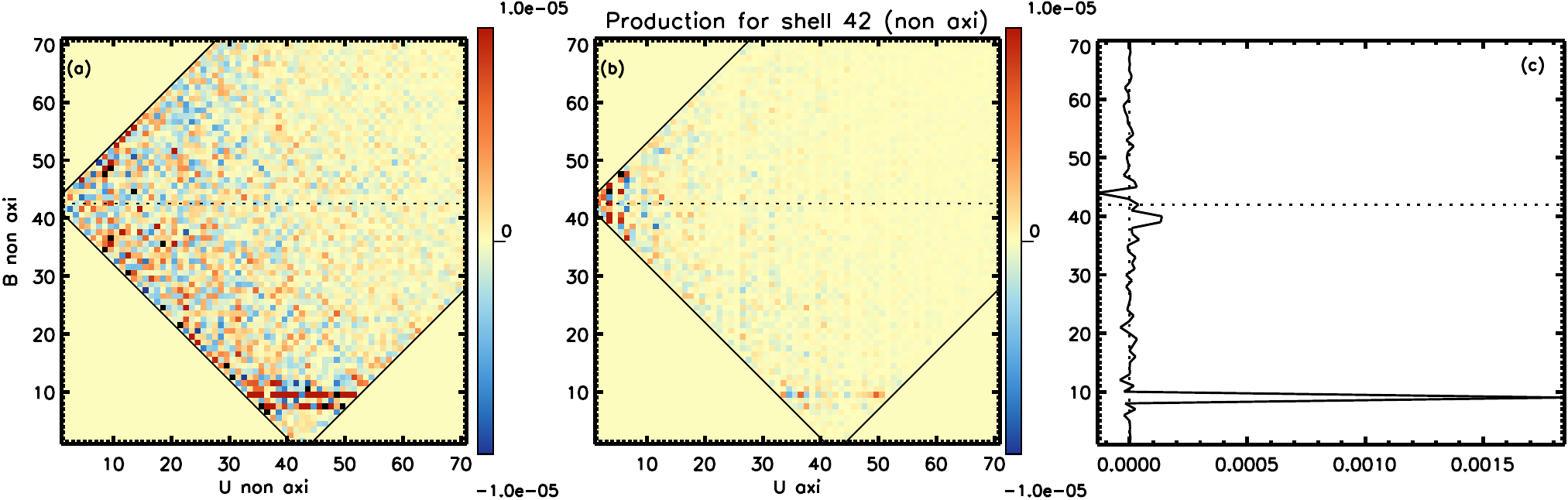}
\caption{Production $\mathcal{P}$ contribution
to the non-axisymmetric shell $L=42$}
  \label{fig:prod_L42_nax_init}
\end{figure*}

\refereee{
The dynamics of zones (I), (III) and (IV) are dominated by two effects
which competes initially: a direct non-local
$\B^\star_9-\U^\star$ transfer of energy, and an effective shearing of
neighbor shells by the large scale differential rotation ($\B^\star -
\U^0_3$ interactions). These two effect are exemplified in
Fig. \ref{fig:prod_L42_nax_init} for $L=42$ (zone
III). 
}

\refereee{
In the case of the large-scale zone (I) \refereee{(not shown here)}, the evolution is dominated by both
$\mathcal{P}$ and $\mathcal{F}$. The interactions between $\B^\star_9$
and $\U^\star$ alternate signs depending on the $\U^\star$ shell
considered. We also \refereee{stress} that the interactions involving other $\B^\star$ shells are not
negligible. The differential rotation action is completely
  negligible compared to $\U^\star-\B^\star$ interactions in zone(I).}

\referee{In the case of the plateau zone (III), almost a flat
profile in the log-log plot is observed in Fig. \ref{fig:evol_ME_sp} (hence its
name). This plateau is characteristic of convective flows that usually exhibit a
broad spectrum between the injection and inertial ranges
(Fig. 
\ref{fig:AB3_hydro}(c)). The evolution of the spectrum is dominated only by the
$\mathcal{P}_L\left(L_1,L_2\right)$ contributions ($\mathcal{F}$ is negligible), and in particular by
the coupling between (non-axisymmetric) $\mathbf{U}^\star$ and $\mathbf{B}^\star$
(Fig. \ref{fig:evol_ME_sum_terms}). Hence, it is a non-local transfer of magnetic
energy that creates the spectrum. \modifff{All the shells in zone (III)
receive energy mainly through this non-local mechanism. As a result, the energy
transfer is very sensitive to the kinetic energy contained in the
$\U_L^\star$ shells involved in the coupling. This explains why the
magnetic energy spectrum reflects the kinetic energy spectrum in this region.} 
}

Although the $\mathbf{U}^\star-\mathbf{B}^\star$ interactions dominate
(Fig. \ref{fig:prod_L42_nax_init}), we stress that
the $\mathbf{U}^0-\mathbf{B}^\star$ interactions exhibit a direct cascade
pattern. $E^\star_{42}$ receives energy from $E^\star_{40}$ through $\mathbf{U}^0_3-\mathbf{B}^\star_{40}$
interactions, and gives energy to $E^\star_{44}$ through $\mathbf{U}^0_3-\mathbf{B}^\star_{44}$
interactions (see panel (b) in
Fig. \ref{fig:prod_L42_nax_init}). Even if the triadic interaction
involves the large scale velocity $\mathbf{U}^0_3$, we nonetheless
refer this effect as a \textit{cascade}. \modif{The velocity field only acts
here as a \textit{mediator}, and the scales of magnetic field
involved in the magnetic energy transfer are at the same scale. It is
consequently a cascade when considering the scales of magnetic field.}  \\

\modiff{The energy transfers in zone (IV) (not shown here) are very
  similar to zone (III). \referee{A noticeable difference is that the
    cascade of energy triggered by the shear of the differential rotation
    $\U^0_3$ is much less efficient since the smallest scales hardly
    feel the large scale rotation profile. Finally, ohmic diffusion acts in the whole zone (IV)
and tends to dissipate energy. It has a sufficiently lower amplitude than the
non-local transfers so that it does not dictate the spectrum shape
initially. It will nevertheless contribute to the saturation process
(Sect. \ref{sec:saturation}).}
}

\subsubsection{Creation of the axisymmetric spectrum}
\label{sec:evol-axisymm-spectr}

\begin{figure}[htbp]
  \centering
  \subfigure[]{
 \includegraphics[width=6cm,angle=90]{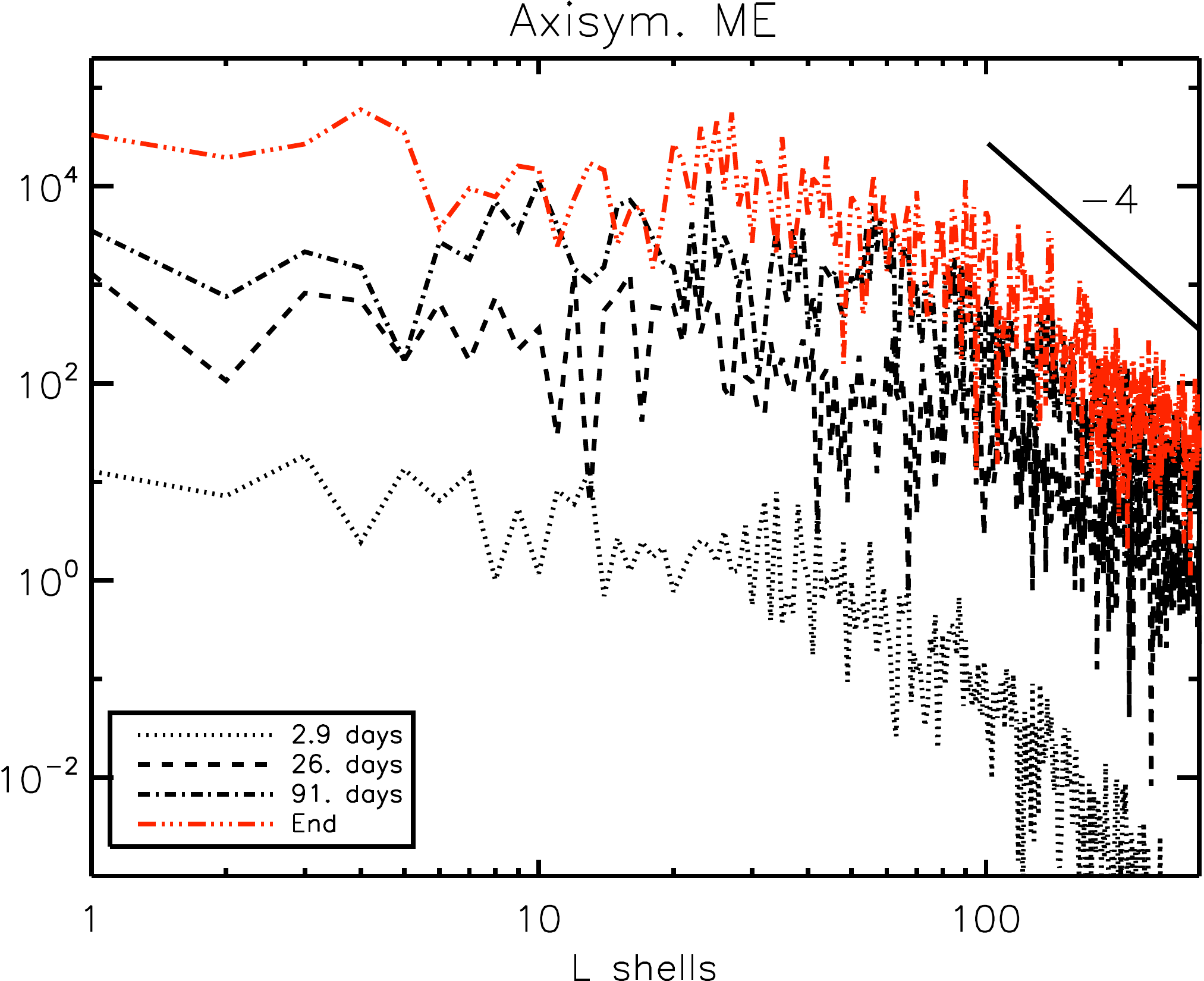}
  \label{fig:evol_ME_axi_sp}
}
  \subfigure[]{
 \includegraphics[width=6cm,angle=90]{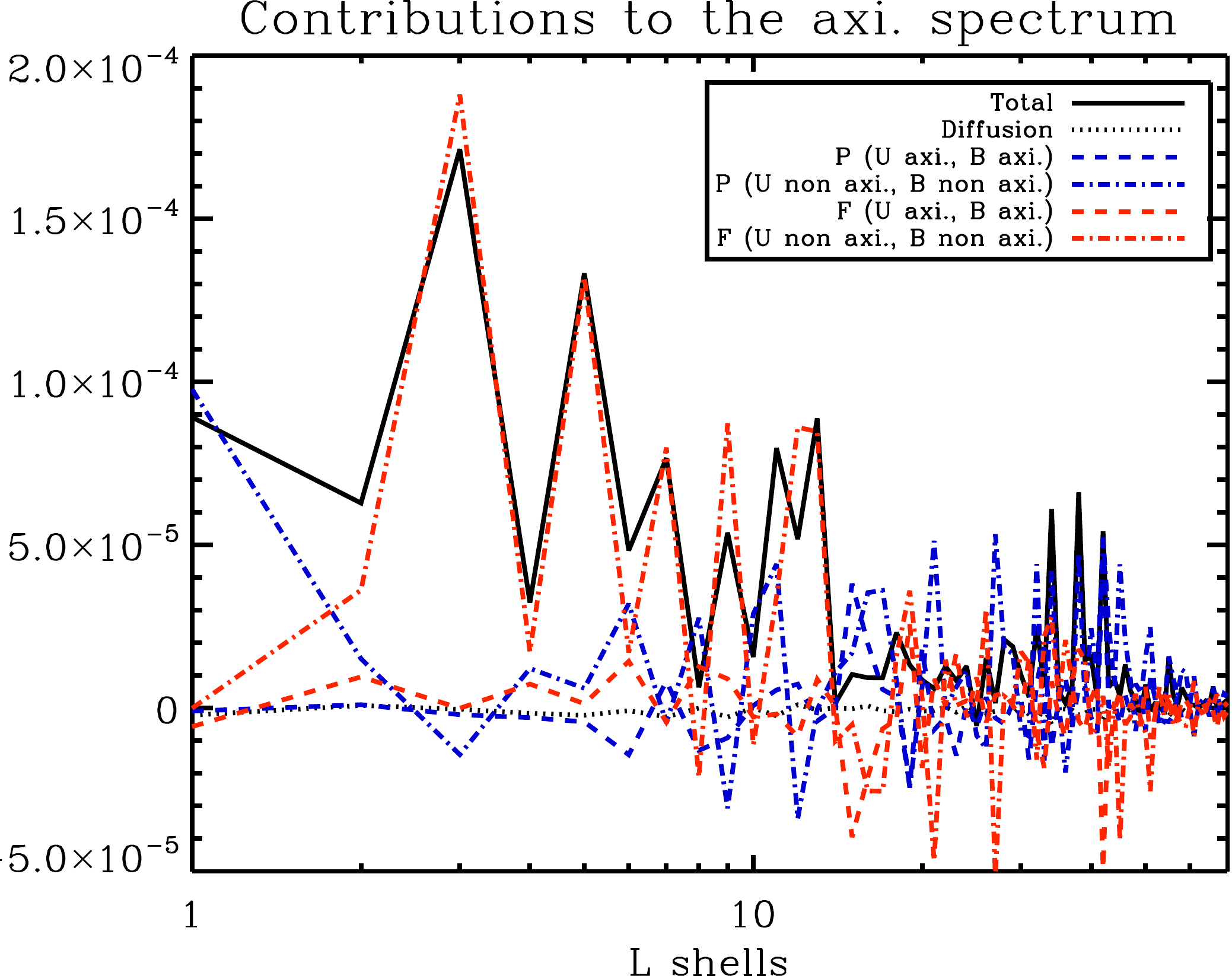}
  \label{fig:evol_ME_axi_sum_terms}
}
  \subfigure[]{
\includegraphics[width=\linewidth]{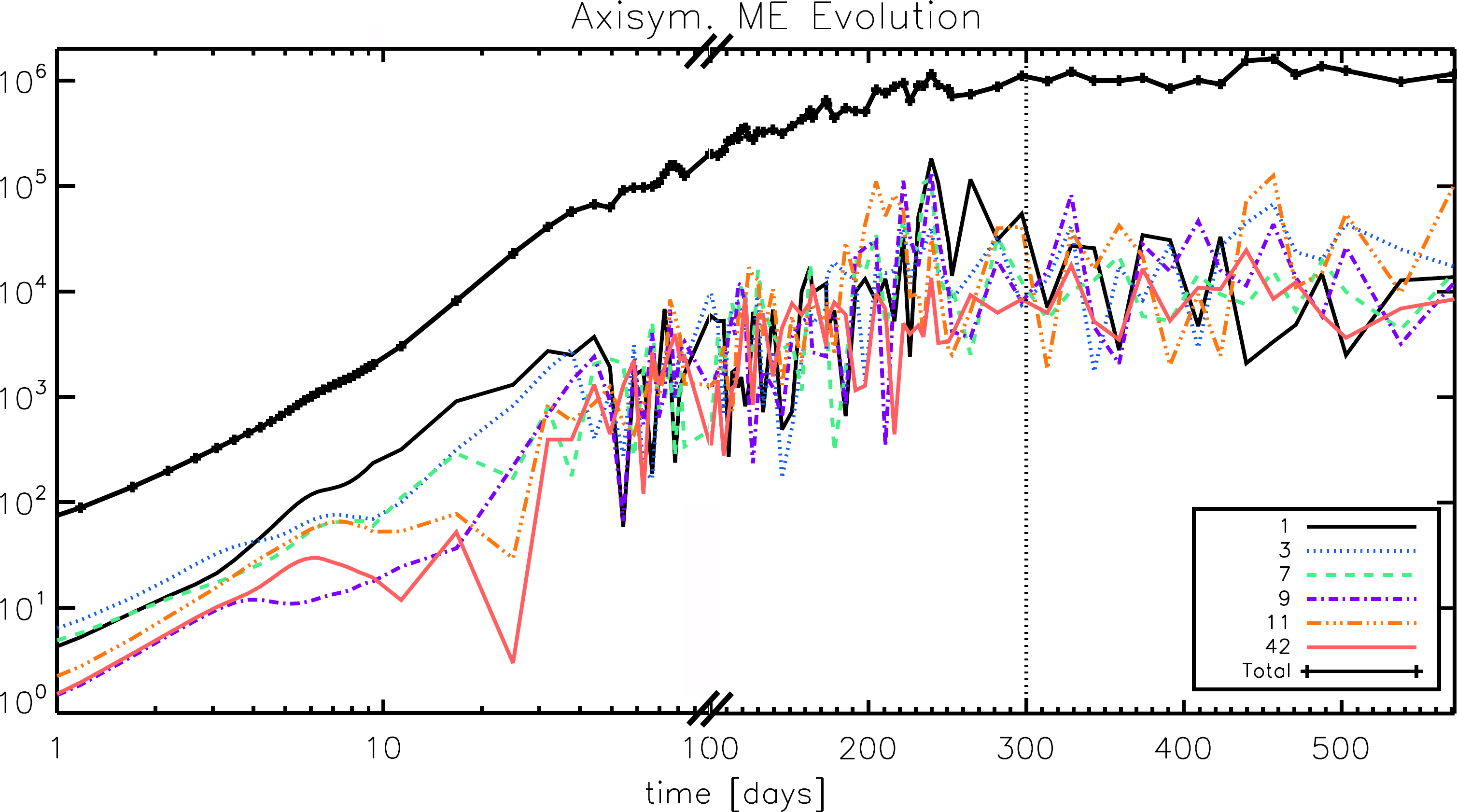}
  \label{fig:evol_shells_axi}
}
  \caption{Same as Fig. \ref{fig:evol_sps_non_axi}, for the
    axisymmetric part of the spectrum. In panel \textbf{(a)}, the
    initial spectrum is zero.}
  \label{fig:evol_sps_axi}
\end{figure}

\begin{figure}[htbp]
  \centering
  \includegraphics[width=8cm]{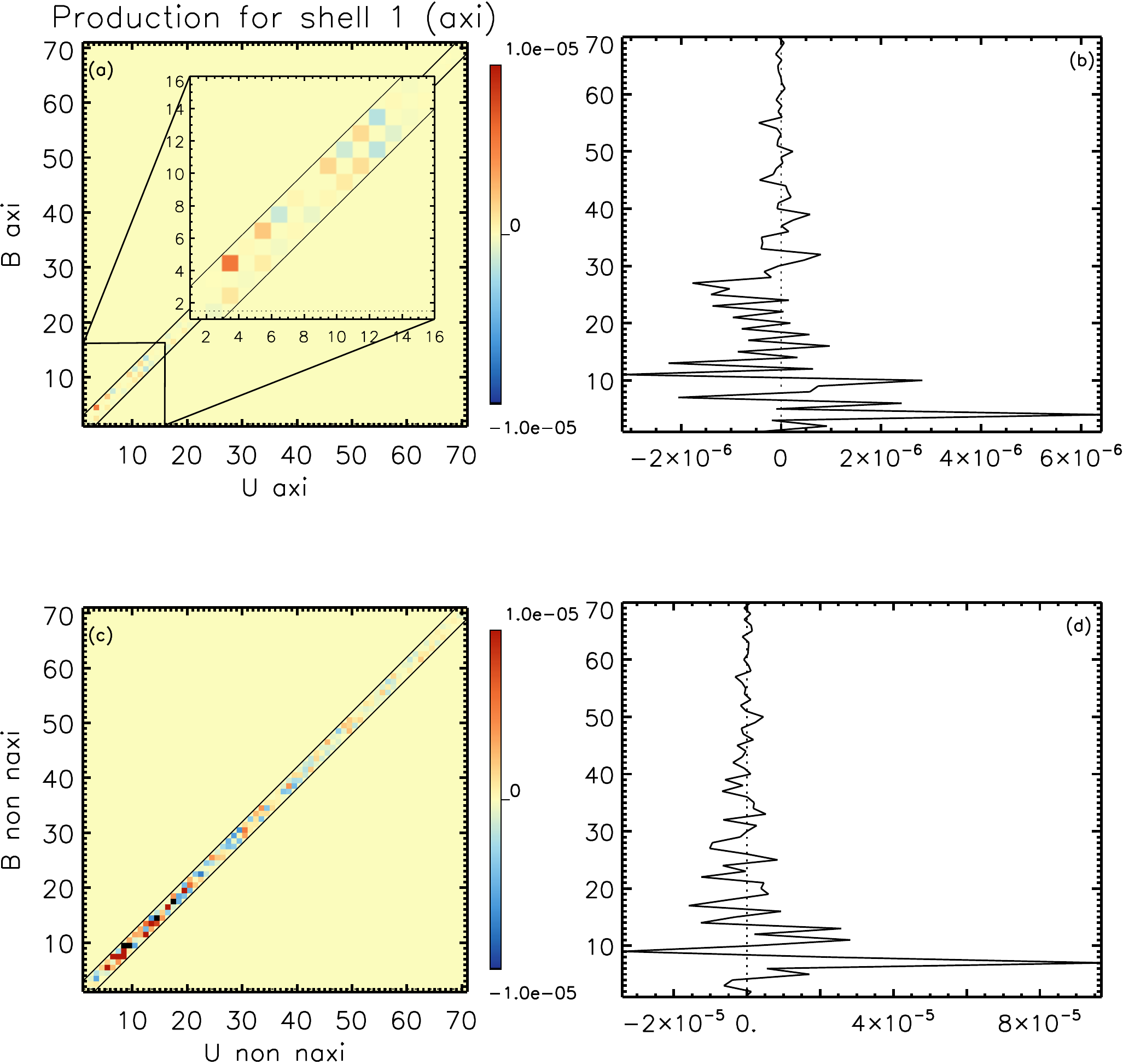}
  \caption{Production $\mathcal{P}$ for the axisymmetric shell
    $L=1$ \modiff{during the initial state}. \modiff{Both $\U^0-\B^0$ and $\U^\star-\B^\star$ couplings are shown.}}
  \label{fig:prod_L1_axi_init}
\end{figure}

We now characterize the creation of the axisymmetric spectrum. \modif{We
  display in Fig. \ref{fig:evol_ME_axi_sp} the evolution of the
axisymmetric component of the magnetic energy. \modifff{We recall that
since we initialize the dynamo with a purely non-axisymmetric field,
the initial axisymmetric spectrum is null. After one time-step, the
axisymmetric magnetic energy is orders of magnitude lower than the
non-axisymmetric spectrum 
(Fig. \ref{fig:AB3_mag}(b)} The global shape of the
axisymmetric spectrum is created very rapidly, all the shells
gain energy at about the same rate until they saturate. \referee{The initial
exponential growth rate is the same for both the axisymmetric and non-axisymmetric
spectrum is approximately $0.6$ days$^{-1}$ (which corresponds to a
time-scale approximately $17$ times lower than the convective
turn-over time).} 
This can also
be observed on Fig. \ref{fig:evol_shells_axi}, where we plot the
evolution of few shells against time. They all gain energy at about
the same rate initially, and then slowly tend to a saturated
state. The axisymmetric shells considered have comparable energy since
the spectrum is essentially flat at scales $L\le 30$ (Fig. \ref{fig:evol_ME_axi_sp}),
which was not the case for the non-axisymmetric spectrum
(Fig. \ref{fig:evol_shells_nax}).}
We observe in Fig. \ref{fig:evol_ME_axi_sum_terms} that the flux term
$\mathcal{F}$ plays a major role between $L=2$ and $L=13$. This means
that the creation of the spectrum is dominated by the radial
interactions at those scales. \modif{The two flux curves exhibit a
  \textit{sawtooth} pattern that is again reminiscent from the differential
rotation energy shells (see Sects. \ref{sec:omega-effect-}
and \ref{sec:evol-non-axisymm})}. At higher
$L$, the evolution of the spectrum is the result of a complex interplay
between the production and flux terms. 

More interesting, the dipole ($L=1$) evolution is dominated by
the production term through the interaction between the non
axisymmetric magnetic field and velocity field. We display the detailed
transfers maps for this scale on Fig. \ref{fig:prod_L1_axi_init}. We
observe that the large scale magnetic field is mainly created by the
interplay between $\B_7^\star$ and $\U^\star_{6-8}$. \modiff{The transfers
involving $\B_9^\star$ (where the energy is originally mainly
contained) \referee{act negatively and} do not dominate the transfer of
magnetic energy. This is consistent with the fact that the whole
axisymmetric spectrum shape is rapidly created and only gains energy
globally afterwards. It does not depend on the scale at which we
initially put the non-axisymmetric magnetic energy. \referee{Since the
energy is not transfered directly from the initial reservoir of energy
$E_9^\star$, we already see preferred transfers towards the large scale
dipole involving $\B^\star_7$, which is one of the highest energy
scale of the non-axisymmetric spectrum at this time.
This effect shall be confirmed during the saturation phase
(Sect. \ref{sec:satur-mean-large}).}
The creation of the axisymmetric magnetic energy
spectrum seems to depend essentially on the initial hydrodynamic
convective spectrum (as expected in such kinematic phase).} \\

\begin{figure*}[htbp]
  \centering
\includegraphics[width=14cm]{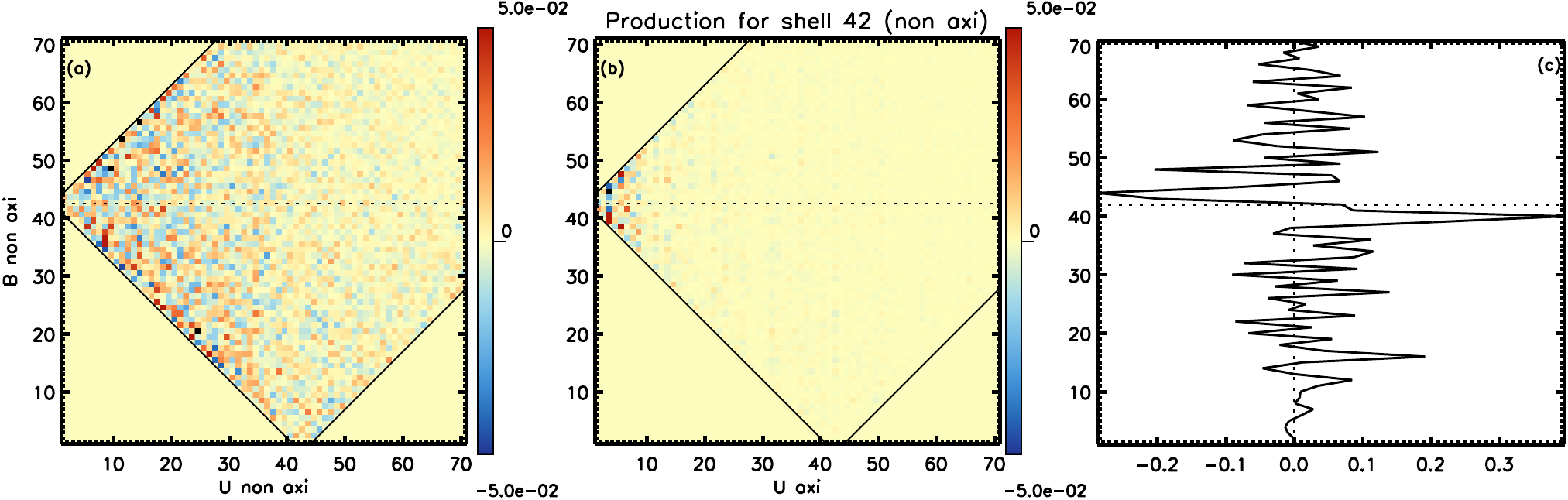}
\caption{Production $\mathcal{P}$ contribution
to the non-axisymmetric shell $L=42$ during the saturation
phase. Interactions between $\B^0$ and $\U^\star$ are negligible.}
  \label{fig:L42_nax_end}
\end{figure*}

\subsection{Non-linear saturation of the smallest scales}
\label{sec:saturation}

Following Sect. \ref{sec:creation-me-spectrum}, we now detail the
saturation and sustainment of the magnetic energy spectrum \referee{at
  small scales}. \refereee{By $500$ days the axisymmetric and non-axisymmetric spectra are
  saturated (Figs. \ref{fig:evol_shells_axi} and
  \ref{fig:evol_shells_nax}).} 

The flux $\mathcal{F}$ contribution is likely
  to never be null at the largest scales since it represents the flux of magnetic energy
  through the horizontal surface at the middle of
  the turbulent convection zone. In order to saturate the magnetic energy \modifff{(\textit{i.e.},
  to get $dE_L^{\rm mag}/dt = 0$)}, 
  $\mathcal{D}$ and/or $\mathcal{P}$ have to compensate
  $\mathcal{F}$. In the first three zones, diffusion is 
  negligible. Hence, $\mathcal{P}$  naturally tends to cancel
  $\mathcal{F}$ out
  in those zones \referee{(see Sect. \ref{sec:mean-field-axisymm} for
    a simple version of this cancellation effect).
The cancellation effect is such that
$\mathcal{F}(\U^\star,\B^\star)$ tends to cancel
$\mathcal{P}(\U^\star,\B^\star)$ out. This is also the case for
$\mathcal{F}(\U^0,\B^\star)$, $\mathcal{F}(\U^\star,\B^0$), and
$\mathcal{F}(\U^0,\B^0)$. }


\modiff{
In spite of the cancellation of the different contributions, characteristic patterns can still be identified. The more
distinctive pattern we identified in
Sect. \ref{sec:creation-me-spectrum} was the direct cascade of
magnetic energy in zone (III). It turns out that we still observe it
and that it slightly dominates the transfer terms during the saturation phase. We
display on Fig. \ref{fig:L42_nax_end} the production
contribution to the non-axisymmetric magnetic energy evolution $600$ days
after the magnetic field introduction. We recover the direct cascade of energy
in the production contribution, that was already present on Fig. \ref{fig:prod_L42_nax_init}. This
direct cascade of energy is associated with an inverse cascade of
energy carried by the flux contribution, which opposes the production
term during the saturation phase. Both cascades are of the same
order of magnitude and tend to cancel each other out. They are
associated with the axisymmetric component $\U^0_3$ (the
differential rotation), and the non-axisymmetric components of
$\B$. \modiffff{The contributions of non-axisymmetric components of $\U$ involve more shells, but
their net effect is a bit lower than the shear from differential
rotation (panel (a) on Fig \ref{fig:L42_nax_end}). On this panel, no
particular global pattern can be identified.}}\\

\begin{figure}[htbp]
  \centering
\includegraphics[width=8.5cm]{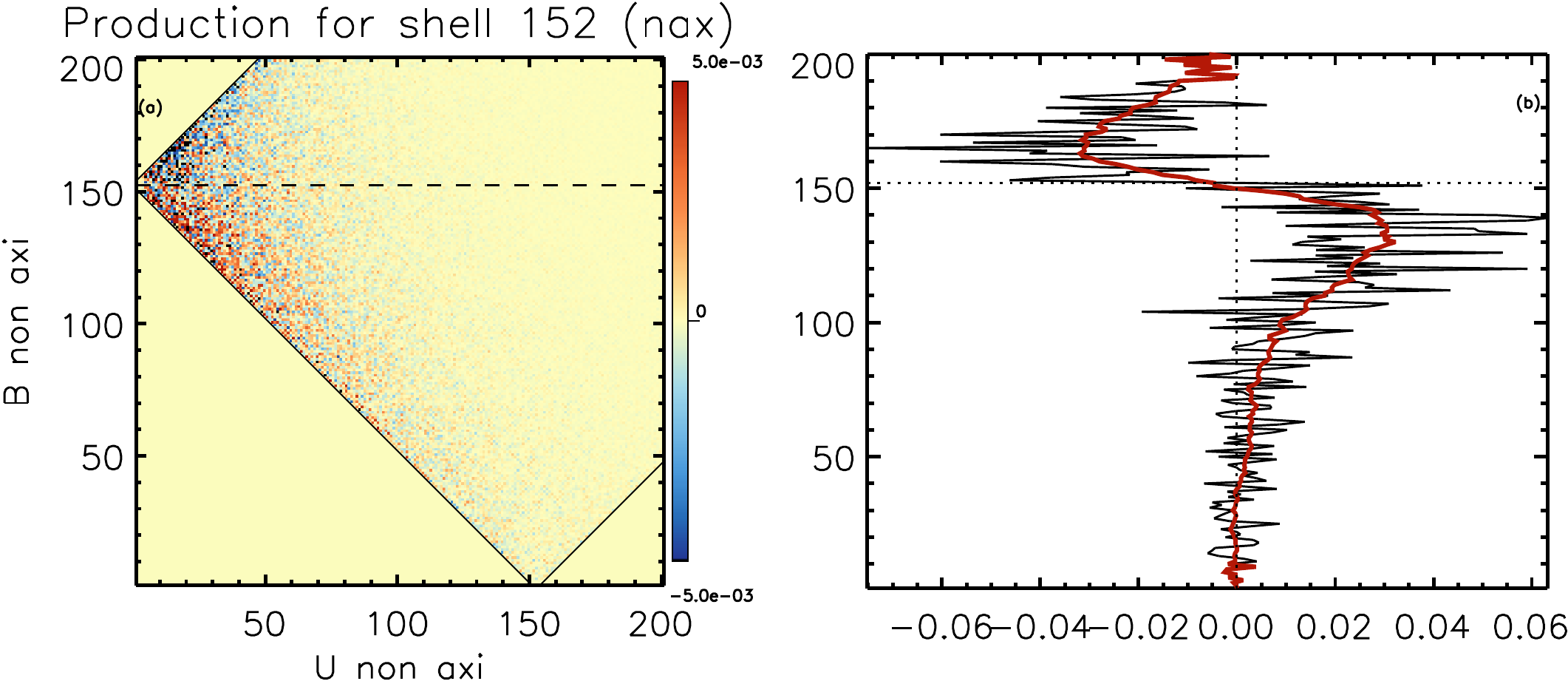}
\caption{Production $\mathcal{P}$ contribution
to the non-axisymmetric shell $L=152$ during the saturation
phase. \referee{The red line is the smoothed contribution, that
  exhibits the characteristic S-shape of the generalized cascade.}}
  \label{fig:L152_nax_end}
\end{figure}

\modiff{\referee{The transfers of magnetic energy appear to be very interesting} in zone (IV) where
  diffusion acts significantly. \modifff{In order to saturate,
    $\mathcal{P}$ and $\mathcal{F}$ have to combine to cancel
    $\mathcal{D}$. For the non-axisymmetric spectrum, it is the
  production term that dominates over the flux term to compensate
  diffusion.} In addition, the production term in zone (IV) exhibits a very particular
generalized cascade shape. This cascade could not be identified during
the early evolution for it was dominated by the non-local transfer
from $\B^\star_9$. We display on Fig. \ref{fig:L152_nax_end}
(panel a) the $\U^\star-\B^\star$ production map towards $E^\star_{152}$. The other
interactions are negligible. We observe that the map is dominated by
positive contribution (red) under the horizontal dashed line
($L=152$), and by negative contribution (blue) above. This is
confirmed by the plot in panel b where the transfers have been summed
over the velocity shells. This cascade is of different kind than the
one observed in zone (III) (Fig. \ref{fig:L42_nax_end}). Here, no clear velocity shell dominates the transfer
map (panel a on Fig. \ref{fig:L152_nax_end}). \referee{ It is a
\textit{generalized cascade} that} results
from the coupling between many magnetic shells (around $L=152$) and
all the largest velocity scales.} \referee{Hence, the velocity scales
involved in the cascade are not local compared to the magnetic field
scale considered.}\\

Trying to simplify the complex 2D transfer maps, one may isolate the main contributing
couplings to the different evolution terms. \referee{Doing so at all times for
the non-axisymmetric spectrum at small scales}, we find that the percentage of couplings that account for
$90$\% of the contributing terms typically varies from nearly $1$ to
$70$\% of the calculated couples. \modifff{As a result, we demonstrate here
that the complex dynamo process occurring in a 3D turbulent convection
zone involves many modes that interact though non-trivial triadic
interactions.} \referee{Then, the dynamics of the smallest scales can hardly be reduced to the
evolution of a small set of modes.}\\

\modiffff{Finally, the analysis of the axisymmetric $\alpha\Omega$
  dynamo in Sect. \ref{sec:mean-field-axisymm} shed light on the
  importance of the families of symmetry (with respect to the equator)
  of the fields. The instantaneous convective motions
  do not exhibit any particular symmetry at any scale and the kinetic
  energy spectrum is a mixture of both primary and secondary
  velocities. The differential rotation is the only velocity feature
  that has a clear symmetry (secondary family, see
  Sect. \ref{sec:omega-effect-}) and that has a
  large influence on the magnetic energy spectrum. It is involved in the magnetic
  energy cascade in zone (III), and shears both primary and secondary
  magnetic fields to cascade primary and secondary magnetic
  energy. Thus, it does not select a particular symmetry. Indeed, the
  ratio of primary (antisymmetric) to secondary (symmetric)
magnetic energy varies with time for all shells and does not settle even during the
saturation phase. The presence of complex flows, often breaking the
equatorial symmetry, yields a strong coupling of both dynamo families
(as in the Sun, see \citet{DeRosa:2012ve}), contrary to simpler mean
field dynamo models (see Sect. \ref{sec:mean-field-kinematic}).
}

\subsection{Sustainment of the mean large scale magnetic field}
\label{sec:satur-mean-large}

\referee{
Given their key role in setting the overall magnetic polarity in the
Sun \citep{DeRosa:2012ve}, we now detail the main contributions to the saturation and sustainment of the
large scale axisymmetric dipole $(l=1,m=0)$ and quadrupole $(l=2,m=0)$ fields.
}

\begin{figure}[htbp]
  \centering
  \includegraphics[width=8cm]{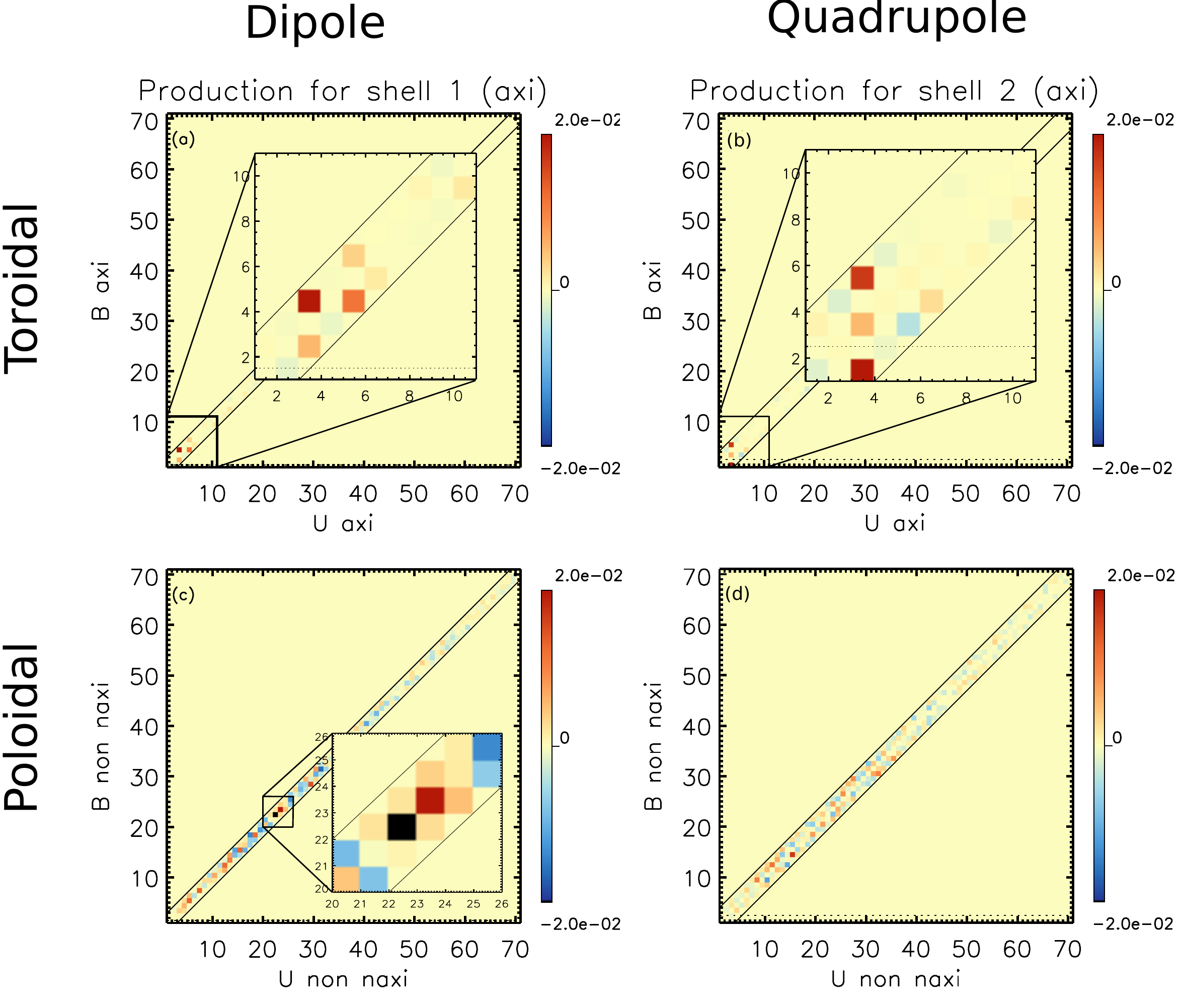}
  \caption{\refereee{Saturation of the large scale axisymmetric dipole
      (panels a and c, production $\mathcal{P}$ for $E^0_1$) and
      quadrupole (panels b and d, production $\mathcal{P}$ for $E^0_2$).
      The interaction maps are time averaged over a period
      of $\sim 150$ days. \modiff{Both $\U^0-\B^0$ and $\U^\star-\B^\star$ couplings are shown.}}}
  \label{fig:large_scale_sat}
\end{figure}

\referee{At the late phase of the simulation the large-scale axisymmetric
  spectrum is fully saturated (Fig. \ref{fig:evol_shells_axi}). The
  saturation is obtained thanks to the compensation of the production
  and flux terms, similarly to the saturation of the mid-scales (see
  previous section). The large-scale dipole $(l=1,m=0)$ saturation
  process differs significantly from its creation. We display in
  Fig. \ref{fig:large_scale_sat} 
  the production maps for the axisymmetric
  dipole \refereee{and quadrupole} averaged over $150$ days during the saturated state. The
  transfers maps of $\mathcal{F}$ \referee{(not shown here)} are exactly opposite to the maps (a)
and (c) \referee{for $\mathcal{P}$}. We see that both the axisymmetric and non-axisymmetric fields
significantly contribute to the saturation and sustainment of the
large scale dipole. In particular, two main contributors emerge. First
(panel a), the coupling of the differential rotation
$\U^0_3$ with the large scale $\B^0_4$ field dominates the
axisymmetric contributions. This effect is more likely to represent
the shearing of the large scale poloidal multipole $\B^0_p$ by the
large-scale toroidal differential rotation.}

\referee{Second, the non-axisymmetric contributions (panel c)
are at least equally important for the saturation of the dipole. In
particular, the interaction $\U^\star_{23}-\B^\star_{23}$ dominates the
non-axisymmetric contributions. Thus, is it a \textit{non-local}
interaction that saturates the large-scale magnetic
dipole. Furthermore, $\B^\star_{23}$ is one of the most energetic shells of the magnetic
energy spectrum (Fig. \ref{fig:evol_ME_sp}). This directly points out
the importance of the mid-scale part of both the kinetic and magnetic
energy spectra for the saturation level of the large-scale magnetic dipole.
}

\referee{We can remark here that the major contributions of
  $\mathcal{P}$ for the saturation of the dipole are all
  positive. They are balanced by negative contributions from
  $\mathcal{F}$. Consequently, if the differential rotation was more
efficient, or if the $\U^\star_{23}-\B^\star_{23}$ interaction possessed more
energy, the saturation level of the large scale dipole would be much
higher.}

\referee{Since our magnetic Prandtl number is $4$, the peak of the
  kinetic and magnetic energy spectra are likely to be shifted. At
  saturation, the couplings are nonetheless dominated by the peak of the
  magnetic energy spectrum that occurs at smaller scale than the peak of
  the kinetic energy spectrum. Changing the magnetic Prandtl number will cause
  the separation of the peaks to change. If the peaks separate more,
  our results suggest that the saturating interaction will involve
  smaller scales velocity and magnetic fields. The velocity field
  involved is likely to be less energetic, which could trigger a
  smaller saturating interaction, and in turn a lower energy state
  for the large scale dipole. If the peaks are closer (or eventually
  switch), the picture becomes more complicated and we cannot predict
  if the saturating interaction will remain fixed by the peak of the
  magnetic energy spectrum. The exploration of this parameter space is
  left for future work.}


\refereee{The large scale quadrupole also saturates thanks to both the
  axisymmetric and non-axisymmetric fields (panels b and d). 
  The axisymmetric
  contributions (panel b) are very similar to the dipole case and are dominated
  by the differential rotation. The differential rotation shears both
  $\B^0_1$ and $\B^0_5$, which is opposed by the flux term to saturate
  the quadrupole. Again, this effect 
  accounts for the saturation of the toroidal quadrupolar
  field. Hence, the saturated level of the poloidal dipolar field
  (panel c) plays
  a major role for the saturation of the toroidal quadrupolar field.}

\refereee{The poloidal quadrupolar field is then saturated through the
  non-axisymmetric interactions (panel d). The contribution are again very
  non-local, though in this case no particular scale dominates the saturation
  process. Hence, we may expect that the saturation process of the
  axisymmetric quadrupole will have a very different dependency on the
  magnetic Prandtl number than
  the axisymmetric dipole. }

\begin{figure*}[htbp]
  \centering
  \includegraphics[width=0.8\linewidth]{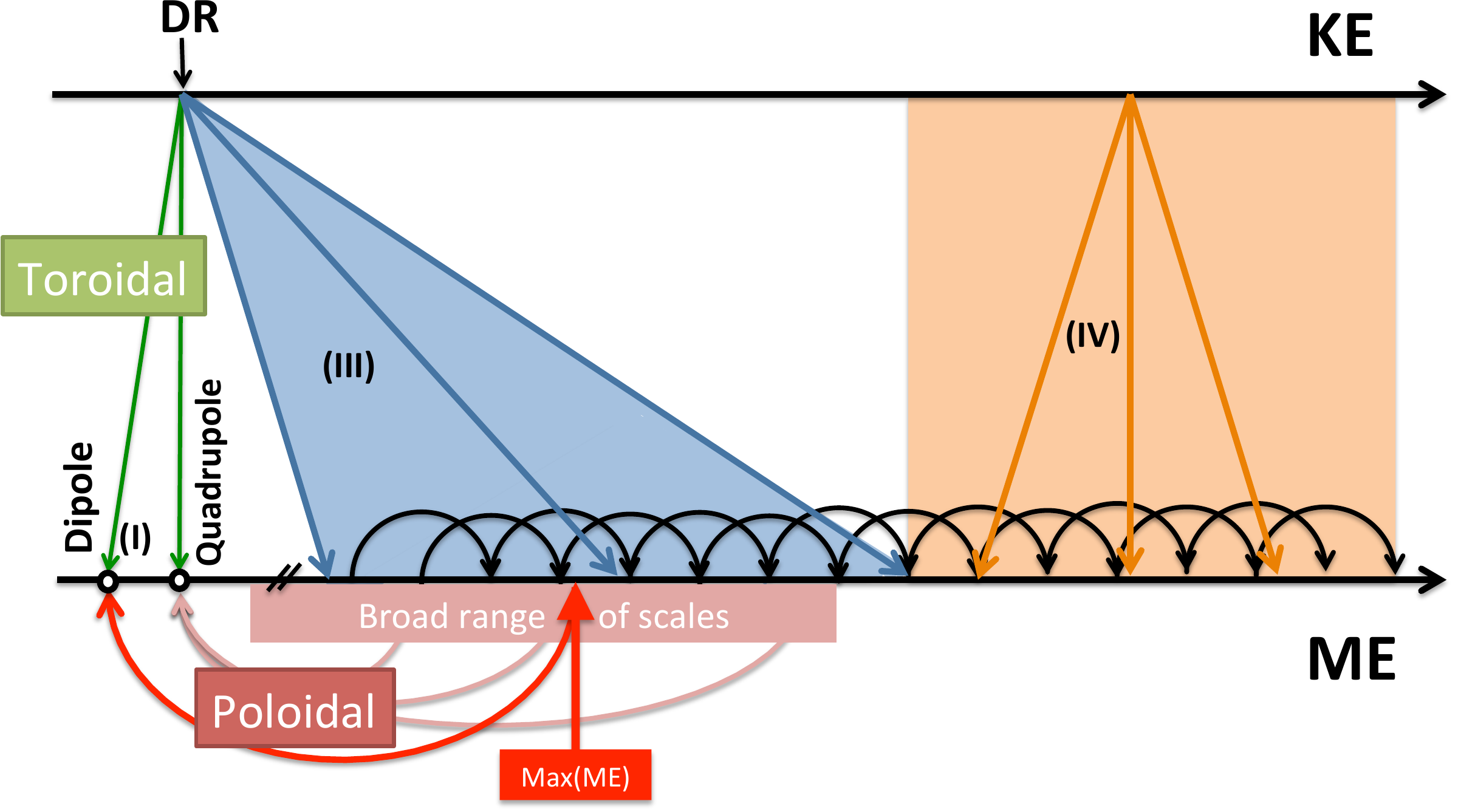}
  \caption{\refereee{Saturating interactions of the non-linear
    convective dynamo summarizing the results of section
    \ref{sec:numer-exper}. KE and ME stand for Kinetic and Magnetic
    Energy and DR for Differential Rotation.}}
  \label{fig:scheme_summary}
\end{figure*}

\section{Conclusions and Perspectives}
\label{sec:concl-persp}

In this paper we developed and validated a new spectral analysis
method suited for spherical objects. Using two vectorial spherical
harmonics basis, we were able to calculate transfer functions of
magnetic energy in spectral space. We can calculate the
coupling coefficients up to $l\sim 500$. For the first time in such
studies, the complete $2$D transfers maps have been calculated to
characterize the full triadic interactions. 

After a quick numerical validation, we first applied our method to a
simplified $\alpha\Omega$ dynamo case. Such axisymmetric models are
very well known to trigger cyclic dynamos
\citep{Charbonneau:2010tw} with our choice of a symmetric (with
respect to equator) velocity field and an antisymmetric $\alpha$
effect. The clear separation between the dipolar and
quadrupolar families was illustrated thanks to our new diagnostic. The
production (\textit{i.e.}, on a spherical surface)
and a flux (\textit{i.e.}, through a spherical surface)
\modif{contributions were shown to quasi-cancel each other out for all
  shells.}
\\

Our method was then successfully applied to a 3D turbulent convective
dynamo case. We initialize a highly non-axisymmetric magnetic field and let
the dynamo develop a turbulent spectrum of magnetic energy. 
\refereee{We distinguished the kinematic phase with exponential growth of the magnetic energy
spectrum, and the non-linearly saturated phase.}
The first phase is
dominated by \referee{both} a non-local transfer of energy from the initial scale of
magnetic energy, coupled with the convective scales, towards all the
other magnetic scales, \referee{and the shearing by the large scale
  differential rotation}. A large part of the magnetic
energy spectrum is then dictated by the kinetic energy spectrum developed
by the convection. 
\\

The \refereee{saturation} phase is more subtle and greatly depends
on the considered scale in the spectrum. \refereee{The saturating
  interactions for the different spectral scales are illustrated in
  Fig. \ref{fig:scheme_summary}, and summarized hereafter. }


Our new method allowed us to
  distinguish two clear cascades of magnetic energy at the smallest
  scales of our simulation, for $13 \le L \le 60$ and $L > 60$
  (highest $L$'s). In the former case, the differential rotation profile mediates the
cascade by shearing the magnetic field. It results in an efficient cascade of 
magnetic energy. 

\referee{
The latter cascade is also direct and involves all the highest
velocity scales (the large scale
differential rotation does not dominate in this case)}. It is a generalized
cascade over a large range of magnetic scales. The velocity
scales involved in the cascade are not local with respect to the
magnetic scales. As consequence, we cannot predict if this generalized
cascade would hold at the lowest scales in the case of a real
convective dynamo where scale separation is much higher. Besides, the
saturation also involve non-local coupling that can eventually be
of the order of the sheared cascade for the intermediate scales. We
proved in that case that the transfers cannot be reduced to a limited
set of modes.
\\

\refereee{The saturation of the large scale axisymmetric dipole and
  quadrupole appear to be radically different than the
  small-scale saturation. The toroidal components are mainly saturated
by the balance of the shearing effect of the differential rotation on the large scale
poloidal fields, and the flux transfer through the spherical shell due
to the effect of the differential rotation.}
The poloidal components are mainly saturated by non-local
non-axisymmetric interactions. The dipole is saturated by the scale of
maximum (highest) magnetic energy, and the quadrupole saturation
is not dominated by any particular scale. These two observations point to
the two main dependencies of the saturating interactions for the
large-scale fields. First, the
rotation rate of the star (which is linked to the saturating
interaction through the differential rotation)
can determine the ability of the dynamo to build wreaths
\citep{Brown:2010cn}, and/or to be in a strong or weak regime
\citep{Christensen:2006jo,Featherstone:2009ft,Simitev:2009ej}. Second,
the magnetic Prandtl number $P_m=\nu/\eta$
determines the postion of the peak of magnetic energy and then affects
the saturating interaction
\citep[\textit{e.g.,} see][]{Schekochihin:2004ip}. We will explore in
detail how the saturating interactions depend on those two effects 
in future work. 

\refereee{Finally, it is worth comparing theses
  results with previous related work of \citet{Livermore:2010dp}. When
using forced helical flows and allowing the dynamo field to back-react
on the flow, they found that the saturation of the large
scale poloidal dipole was dominated by non-local interaction with a particular
magnetic scale ($l=10,m=10$ in their case). They showed that magnetic
energy was transfered to this
scale by the large-scale toroidal magnetic field. In our case, the large-scale dipole is also saturated
due to non-local interactions. Though, the flow we consider is
significantly different because \textit{(i)} it is obtained from the
convective instability and \textit{(ii)} its spectrum is dominated by
the large-scale differential rotation that develops self-consistently.
Hence, the dynamo process is different and we find that the
toroidal large-scale field is saturated by the effect of the large
scale differential
rotation, and the large-scale poloidal dipole by the non-local transfer of
energy from the magnetic scale of maximal energy.}
\\

\referee{Our results also suggest that no significant large-scale
  magnetic field is growing over dissipative time scales in our
  simulation (the ohmic dissipation time scale for the axisymmetric
  dipole is typically of the order of $\tau_\eta\sim 700$ days in the
  simulation). Again, the fast saturation of the dynamo (less than
  $300$ days) may not hold
  for lower magnetic Prandtl number dynamos.}\\

\referee{We developed a diagnostic on
the magnetic energy that is an invariant of ideal MHD. In the case of
non-ideal MHD, the existence of the selective decay
\citep{Taylor:1974bj,Matthaeus:1980eo,Mininni:2006fd} introduces a decoupling
between, \textit{e.g.}, the evolution
time scales of the magnetic (or total) energy (fast) and the magnetic
helicity (slow). As mentioned before, we were interested, in this work, in fast phenomena
compared to the ohmic diffusion time. For such processes, the ideal invariants
of MHD are still the appropriate quantities to
interpret the scales interactions. \refereee{The dynamo saturation is
  necessarily achieved through a modification of the kinetic energy
  spectrum. As mentioned before, the case we studied in this paper is in the weak branch of
the dynamo \citep[\textit{i.e.}, the large scale poloidal magnetic field does
not dominate the magnetic energy
spectrum][]{Christensen:2006jo,Simitev:2009ej,Gastine:2012kl}.
The detailed modification of the kinetic energy spectrum implied by the saturation
of the dynamo process will be discussed for both the strong and weak
branches in a future work.} Although the magnetic energy is the
relevant quantity to characterize
nearly kinematic dynamos (where the Lorentz force plays little role),
a diagnostic on kinetic energy \refereee{would be highly valuable for
  non-linearly saturated dynamos.} 
On top of that, the detailed spectral
transfers of magnetic helicity are also mandatory to fully address the
complexity of the dynamo process. Evolution equations of kinetic energy,
magnetic helicity and cross helicity in the framework introduced in this paper are
under development and will be published in a forthcoming paper.} \\

 Finally, those diagnostics may also prove very useful for non-dynamo
 transfers related MHD 
phenomena. \modiff{For example, spectral
 analysis applied to the relaxation and the stability of low-l fossil
  field \citep[see][]{Braithwaite:2006da,Brun:2007dd,Zahn:2007ka,Duez:2010hg,Duez:2010jp} will be studied in a
  future publication.}

\acknowledgments

We acknowledge inspiring interactions with the participants
of the fifth and sixth Festival de Th\'eorie, held in Aix en Provence,
July 2009 and 2011. \referee{We acknowledge the valuable comments of the
anonymous referee that have tightened the focus of this paper.}
A. Strugarek acknowledges S. Matt for his careful reading of the manuscript.
A. S. Brun and A. Strugarek acknowledge funding
by the European Research Council through ERC grant STARS2 207430
(www.stars2.eu). The simulations were performed using HPC resources of
GENCI [IDRIS], project 1623.

\bibliographystyle{apj}
\bibliography{bib_astro2}

\appendix
\section{Definition and properties of Vectorial Spherical Harmonics}
\label{sec:defin-prop-vect}

\subsection{Classical vectorial spherical harmonics basis}
\label{sec:leftrlm-slm-tlmright}

\subsubsection{Definitions}
\label{sec:definitions}

We define from \citet{Rieutord:1987go,Mathis:2005kz}:
\begin{equation}
  \label{eq:RST_annex}
  \left\{
  \begin{array}{lcl}
    \Rlm{}(\theta,\varphi) &=& \Ylm{}(\theta,\varphi) \er \\
    \Slm{}(\theta,\varphi) &=& \gradperp \Ylm{} = \dth\Ylm{}\etheta + \frac{1}{\sin{\theta}}\dphi\Ylm{}\ephi\\
    \Tlm{}(\theta,\varphi) &=& \gradperp\times\Rlm{} = \frac{1}{\sin{\theta}}\dphi\Ylm{}\etheta  -\dth\Ylm{}\ephi
  \end{array}
  \right. ,
\end{equation}
where $(\mathbf{e}_r,\mathbf{e}_\theta,\mathbf{e}_\varphi)$ defines the
spherical basis and $Y^m_l$ are the Laplace spherical harmonics
defined by
\begin{equation}
  \label{eq:norm}
  Y_l^m(\theta,\varphi) =
  (-1)^{\frac{m+|m|}{2}}\sqrt{\frac{(2l+1)}{4\pi}\frac{(l-|m|)!}{(l+|m|)!}}P^{|m|}_l(\cos
  \theta)e^{im\varphi}\, 
\end{equation}
\modiff{where $P^m_l$ are the associated Legendre polynomials.}
The basis \eqref{eq:RST_annex} have the following properties :
\begin{eqnarray}
  \label{eq:R.R}
  \int_S \Rlm{1} \cdot \left( \Rlm{2} \right)^{cc} \dint{\Omega}{} &=&
  \delta_{l_1,l_2}\delta_{m_1,m_2}, \\
  \label{eq:S.S_T.T}
  \int_S \Slm{1} \cdot \left( \Slm{2} \right)^{cc} \dint{\Omega}{} &=&
  \int_S \Tlm{1} \cdot \left( \Tlm{2} \right)^{cc} \dint{\Omega}{} =
  l_1(l_1+1)\delta_{l_1,l_2}\delta_{m_1,m_2} ,
\end{eqnarray}
where $S$ is a spherical surface, $\dint{\Omega}{}=\sin\theta \mbox{d}\theta\mbox{d}\varphi$ the solid
angle, $cc$ means complex conjugate and $\delta$ is the Kronecker symbol. We also have:
\begin{equation}
  (\Slm{})^{cc} = (-1)^m\mathbf{S}^{-m}_l \, ,
  \label{eq:conjg_rst}
\end{equation}
and all the other scalar cross products are $0$. We remind
the reader that the poloidal fields are described by their projection
on $(\mathbf{R}^m_l,\mathbf{S}^m_l)$, and the toroidal fields by their
projection on $\mathbf{T}^m_l$.

\subsubsection{Scalar fields}
\label{sec:scalar-fields}

Defining $\psi(r,\theta,\varphi)=\sumlm{}\left\{\psi_m^l(r)Y_l^m(\theta,\varphi)\right\}$, we get:
\begin{eqnarray}
  \label{eq:gradient}
  \grad\psi&=& \sumlm{} \left\{ \dr\psi_m^l\Rlm{} +
    \frac{\psi_m^l}{r}\Slm{} \right\},\\
  \label{eq:lapla}
  \Div\grad\psi &=& \sumlm{} \Delta_l\psi_m^lY_l^m
\end{eqnarray}
where $\Delta_l=\drr + \frac{2}{r}\dr-\frac{l(l+1)}{r^2}$. 

\subsubsection{Vectorial fields}
\label{sec:vectorial-fields}

For a vector $\mathbf{X}(r,\theta,\varphi)=\sumlm{}\left\{\mathcal{A}^l_m(r)\Rlm{} + \mathcal{B}^l_m(r)\Slm{}
  + \mathcal{C}^l_m(r)\Tlm{}\right\}$, we obtain:
  \begin{eqnarray}
    \label{eq:divergence}
    \Div\mathbf{X} &=&
    \sumlm{}\left\{\left[\frac{1}{r^2}\dr(r^2\MAlm{})-l(l+1)\frac{\MBlm{}}{r}\right]Y_l^m\right\}, \\
    \label{eq:curl}
    \rot\mathbf{X} &=&
    \sumlm{}\left\{\left[l(l+1)\frac{\MClm{}}{r}\right]\Rlm{} +
    \left[\frac{1}{r}\dr(r\,\MClm{})\right]\Slm{}  +
    \left[\frac{\MAlm{}}{r}-\frac{1}{r}\dr(r\, \MBlm{})\right]\Tlm{}\right\}, \\
    \label{eq:lapla_vect}
    \nabla^2\mathbf{X} &=&
    \sumlm{}\left\{\left[\Delta_l\MAlm{}-\frac{2}{r^2}(\MAlm{}-l(l+1)\MBlm{})\right]\Rlm{}\right.
    \nonumber \\
    &+&\left. \left[\Delta_l\MBlm{}+2\frac{\MAlm{}}{r^2}\right]\Slm{}  +
    \left[\Delta_l\,\MClm{}\right]\Tlm{}\right\}\, .
  \end{eqnarray}

\subsubsection{Recurrence relations}
\label{sec:recurrence-relations}

In addition to the expression of the different operators, we also give
here two useful coupling relations between spherical harmonics. First,
according to \citet{Varshalovich:1988ul}, 
the coupling between $\cos{\theta}$ and the spherical harmonic $Y_l^m$
is given by
\begin{equation}
  \label{eq:costheta_ylm}
  \cos{\theta}\, Y_l^m =
  \sqrt{\frac{\left(l-m+1\right)\left(l+m+1\right)}{\left(2l+1\right)\left(2l+3\right)}}Y^m_{l+1}
  +
  \sqrt{\frac{\left(l-m\right)\left(l+m\right)}{\left(2l-1\right)\left(2l+1\right)}}Y^m_{l-1}
  \, .
\end{equation}
Then, one easily deduces the following properties:
\begin{eqnarray}
  \cos{\theta}\, \Slm{} &=&
  \frac{l}{l+1}\sqrt{\frac{\left(l-m+1\right)\left(l+m+1\right)}{\left(2l+1\right)\left(2l+3\right)}}
  \mathbf{S}^m_{l+1} +
  \frac{l+1}{l}\sqrt{\frac{\left(l-m\right)\left(l+m\right)}{\left(2l-1\right)\left(2l+1\right)}}
  \mathbf{S}^m_{l-1} \nonumber \\
&-& \frac{im}{l\left(l+1\right)}\Tlm{} \, ,
  \label{eq:costheta_slm}
\\
  \cos{\theta}\, \Tlm{} &=&
  \frac{l}{l+1}\sqrt{\frac{\left(l-m+1\right)\left(l+m+1\right)}{\left(2l+1\right)\left(2l+3\right)}}
  \mathbf{T}^m_{l+1} +
  \frac{l+1}{l}\sqrt{\frac{\left(l-m\right)\left(l+m\right)}{\left(2l-1\right)\left(2l+1\right)}}
  \mathbf{T}^m_{l-1} \nonumber \\
  &+& \frac{im}{l\left(l+1\right)}\Slm{} \, .
  \label{eq:costheta_tlm}
\end{eqnarray}

\subsection{An alternative vectorial basis}
\label{sec:mathbfym_l-l+nu-basi}

\subsubsection{Definitions}
\label{sec:definitions-1}

The vectorial spherical harmonics basis defined in appendix
\ref{sec:leftrlm-slm-tlmright} is very efficient to calculate scalar
products or linear differential operator on vectors. Nevertheless, it
is quite hard to use it to express vectorial products. Instead we
define the following basis (\textit{e.g.}, see \citet{Varshalovich:1988ul}): 
\begin{equation}
  \label{eq:base2_annex}
  \Ylmn{}(\theta,\varphi) =  \sum_{\mu=-1}^{1} \left\{ (-1)^{l-m}\sqrt{2l+1}
  \left(
 \begin{array}{ccc}
    l & l+\nu & 1 \\
    m & \mu-m & -\mu
  \end{array}
  \right)
  Y_{l+\nu}^{m-\mu} \mathbf{e}_\mu \right\},
\end{equation}
where $(\dots)$ is the $3$-j Wigner coefficient linked to Clebsch-Gordan
coefficients, and the vectors $\mathbf{e}_\mu$ are
\begin{equation}
  \label{eq:emu}
  \left\{
  \begin{array}{lcl}
    \mathbf{e}_{-1} &=& \frac{1}{\sqrt{2}}\left(\mathbf{e}_x
      -i\mathbf{e}_y \right) \\
    \mathbf{e}_{0} &=& \mathbf{e}_z \\
    \mathbf{e}_{1} &=& - \frac{1}{\sqrt{2}}\left(\mathbf{e}_x
      +i\mathbf{e}_y \right)
  \end{array}
  \right. ,
\end{equation}
where $(\mathbf{e}_x,\mathbf{e}_y,\mathbf{e}_z)$ defines
the cartesian basis.
Note that the equivalent of the conjugation rule \eqref{eq:conjg_rst}
is then 
\begin{equation}
  \label{eq:conjg_Ylmn}
  \left(\mathbf{Y}^m_{l,l+\nu}\right)^{cc} =
  \left(-1\right)^{m+\delta_{0\nu}} \mathbf{Y}^{-m}_{l,l+\nu}\, .
\end{equation}
Again, we recall that the poloidal
fields are described by their projection on
$(\mathbf{Y}^m_{l,l+1},\mathbf{Y}^m_{l,l-1})$ ($\nu \in
\left\{-1;1\right\}$), and the toroidal fields are described by their
projection on $\mathbf{Y}^m_{l,l}$ ($\nu=0$).

\subsubsection{Vectorial product}
\label{sec:vectorial-prodcut}

We decompose a vector $\mathbf{X}$ on this basis in the following way:
\begin{eqnarray*}
 \mathbf{X}(r,\theta,\varphi) = \sumlmn{} X_{l,l+\nu}^{\hspace{0.1cm}m}(r)\Ylmn{}.
\end{eqnarray*}
Evaluating the vectorial product of two vectors
$\mathbf{X}_{12} = \mathbf{X}_1\times \mathbf{X}_2$, one gets:
\begin{equation}
  \label{eq:vect_prod}
  \Xlmn{12} = \mysum{1}{2}{12} \Xlmn{1}\Xlmn{2}
  \mathcal{J}^{l_{12},m_{12},\nu_{12}}_{l_1,m_1,\nu_1,l_2,m_2,\nu_2} ,
\end{equation}
where 
\begin{eqnarray}
  &&\mathcal{J}^{l,m_1+m_2,\nu}_{l_1,m_1,\nu_1,l_2,m_2,\nu_2} =
  i(-1)^{\nu_1-\nu_2+(m_1+m_2)}\nonumber
  \\
  &&\sqrt{\frac{3}{2\pi}(2l_1+1)(2l_1+2\nu_1+1)(2l_2+1)(2l_2+2\nu_2+1)(2l+1)(2l+2\nu+1)}\nonumber
  \\
  &&
  \left\{
 \begin{array}{ccc}
    l_1 & l_2 & l \\
    l_1+\nu_1 & l_2+\nu_2 & l+\nu \\
    1 & 1 & 1
  \end{array}
  \right\}
  \left(
 \begin{array}{ccc}
    l_1 & l_2 & l \\
    m_1 & m_2 & -(m_1+m_2)
\end{array}
  \right)
  \left(
 \begin{array}{ccc}
   l_1+\nu_1 & l_2+\nu_2 & l+\nu  \\
   0 & 0 & 0
\end{array}
  \right),
  \label{eq:J_clebsch}
\end{eqnarray}
with $\{\cdots\}$ being the $9$-j Wigner coefficient.

\subsection{Basis change relations}
\label{sec:base-change-relat}

For a vector $\mathbf{X}$ decomposed in the following manner:
\begin{eqnarray*}
 \mathbf{X} &=& \sumlm{} \left\{\mathcal{A}^l_m\Rlm{} + \mathcal{B}^l_m\Slm{}
  + \mathcal{C}^l_m\Tlm{} \right\} \\
  &=& \sumlmn{} \left\{X_{l,l+\nu}^{\hspace{0.1cm}m}\Ylmn{}\right\},
\end{eqnarray*}
we have the two following relations to change from one basis to the other:
\begin{equation}
  \label{eq:base_change}
  \left\{
    \begin{array}{lcl}
      \mathcal{A}^l_m &=& \frac{1}{\sqrt{2l+1}}\left[
        \sqrt{l}X_{l,l-1}^{\hspace{0.1cm}m} -
        \sqrt{l+1}X_{l,l+1}^{\hspace{0.1cm}m} \right] \\
     \mathcal{B}^l_m &=& \frac{1}{\sqrt{2l+1}}\left[
        \frac{1}{\sqrt{l}}X_{l,l-1}^{\hspace{0.1cm}m} +
        \frac{1}{\sqrt{l+1}}X_{l,l+1}^{\hspace{0.1cm}m} \right] \\
     \mathcal{C}^l_m &=& \frac{i}{\sqrt{l(l+1)}}X_{l,l}^{\hspace{0.1cm}m}
   \end{array}
  \right. \mbox{ ,}
  \left\{
    \begin{array}{lcl}
      X_{l,l-1}^{\hspace{0.1cm}m} &=& \sqrt{\frac{l}{2l+1}}\left( \mathcal{A}^l_m +
        (l+1) \mathcal{B}^l_m\right) \\
      X_{l,l}^{\hspace{0.1cm}m} &=& -i\sqrt{l(l+1)} \mathcal{C}^l_m \\
      X_{l,l+1}^{\hspace{0.1cm}m} &=& \sqrt{\frac{l+1}{2l+1}}\left( - \mathcal{A}^l_m +
        l \mathcal{B}^l_m\right)
    \end{array}
  \right. .
\end{equation}

\subsection{Expression of the $\alpha$ effect}
\label{sec:expr-alpha-effect}
\modif{
The $\alpha$ effect introduces the spectral coupling of a scalar and a
vector, which was not treated before. In the special case of an
axisymmetric $\alpha$ and an axisymmetric vectorial field $B_\varphi\mathbf{e}_\varphi$ (which
is the case in this paper, see Sect. \ref{sec:mean-field-kinematic}),
we write the $\alpha$ coefficient
\begin{equation*}
  \alpha(r,\theta) = \sum_{l=0}^{+\infty} \alpha^l_0(r) Y^0_l \, ,
\end{equation*}
and we rewrite the magnetic field from \eqref{eq:B2}
\begin{equation*}
  B_\varphi\mathbf{e}_\varphi = \sum_{l=0}^{+\infty}
  \frac{A^l_0}{r}\mathbf{T}^0_l \, .
\end{equation*}
Introducing the coefficient
\begin{equation}
  \label{eq:Hcoef_alphaeffect}
  \mathcal{H}^{l}_{l_1,l_2} = -\sqrt{\frac{1}{4\pi}l(l+1)l_2(l_2+1)(2l+1)(2l_1+1)(2l_2+1)}
  \left(
 \begin{array}{ccc}
    l_1 & l_2 & l \\
    0 & 0 & 0
\end{array}
  \right)
  \left(
 \begin{array}{ccc}
   l_1 & l_2 & l  \\
   0 & 1 & -1
\end{array}
  \right)\, ,
\end{equation}
we can write the $\alpha$ effect such as
\begin{equation}
  \label{eq:alpha_eff_spectr}
   \alpha B_\varphi\mathbf{e}_\varphi
     = \sum_{l=0}^{+\infty} \left( \sum_{\substack{l_{1},l_{2} = 0 \\ l \ge |l_{1}-l_{2}| \\
      l \le l_1 + l_2}  }^{\infty} \mathcal{H}^{l}_{l_1,l_2}
  \alpha^0_{l_1} \frac{A^0_{l_2}}{r} \right) \,\, \mathbf{T}^0_l \equiv
\sum_{l=0}^{+\infty} \, \frac{\xi^l_0}{r} \, \mathbf{T}^0_l \, .
\end{equation}
Finally, one gets
\begin{equation}
  \label{eq:alpha_total_eff_spectr}
   \bnab\times\left(\alpha B_\varphi\mathbf{e}_\varphi\right) =
   \sum_{l=0}^{+\infty} \left\{  \frac{l(l+1)}{r^2}\xi^l_0
   \mathbf{R}^0_l + \frac{1}{r}\partial_r\left(\xi^l_0\right)
   \mathbf{S}^0_l \right\} \, .
\end{equation}
}
\subsection{Couplings in the magnetic energy equation}
\label{sec:couplings-mag}

The detailed expressions of the different terms of the magnetic energy
equation \eqref{eq:Diff_term}-\eqref{eq:Flux_term} are given here.
We write the magnetic field $\B$ and the current $\J$
\begin{align}
  \label{eq:B_J_cleb}
  \B &= \sum_{l=1}^{+\infty}\sum_{m=-l}^{l} \sum_{\nu=-1}^{1}
  B^m_{l,l+\nu}(r) \Ylmn{}\, , \\
  \J &= \sum_{l=1}^{+\infty}\sum_{m=-l}^{l} \sum_{\nu=-1}^{1}
  J^m_{l,l+\nu}(r) \Ylmn{} \, .
\end{align}
In this basis, the vectorial product may be evaluated thanks to a coupling coefficient $\mathcal{J}^{l,m,\nu}_{l_1,m_1,\nu_1,l_2,m_2,\nu_2}$ given
in equation \eqref{eq:J_clebsch}. The transformation rules from one
basis to the other are given in appendix \ref{sec:base-change-relat},
and they allow us to easily evaluate the integrals
\eqref{eq:Diff_term}-\eqref{eq:Flux_term}. 
By separating the diffusive terms into two $\mathcal{D}_1$ and
$\mathcal{D}_2$ terms, we get that
\begin{align}
    \mathcal{D}_1(L,r) &= \eta  \sum_{L} (-1)^m l(l+1) \left\{
    \frac{l(l+1)}{r^3}\Delta_l\left(\frac{\Clm{}}{r}\right)C_{-m}^l \right.
    +
    \frac{1}{r^2}\dr\left[r\Delta_l\left(\frac{\Clm{}}{r}\right)\right]\dr
      C_{-m}^l 
      + \Delta_l\left(\frac{\Alm{}}{r}\right) \left.\frac{A_{-m}^l}{r} \right\},
  \label{eq:diff1_full_expr}
  \\
  \label{eq:diff2_full_expr}
  \mathcal{D}_2(L,r) &=
  -\frac{c\,\dr\eta}{\sqrt{4\pi}} \sum_{L}\sum_{\nu_1,\nu_2}
  B_{l,l+\nu_1}^{\hspace{0.1cm}m} J_{l,l+\nu_2}^{-m} 
\mathcal{J}^{0,0,1}_{l,m,\nu_1,l,-m,\nu_2}\, ,
\end{align}
where $\sum_L$ stands for a summation over all the spherical harmonics
contained in the shell $L$ (one element in an axisymmetric shell, and
$2l$ elements in a non-axisymmetric shell). The production and flux terms then read
\begin{align}
  \label{eq:prod_full_expr}
  \mathcal{P}_L\left(L_1,L_2,r\right) &= 
\frac{c}{4\pi}\sum_{L} \frac{(-1)^m}{r^2} \left.\frac{l(l+1)}{\sqrt{2l+1}}
   \right\{ 
   \left[\sqrt{l} \right. \Klm{(\U_{L_1}\times\B_{L_2})}{}{-1}
   \left. -
     \sqrt{l+1} \Klm{(\U_{L_1}\times\B_{L_2})}{}{+1} \right] A_{-m}^l
   \nonumber \\
   &+
 \left[\frac{1}{\sqrt{l}}\right. \Klm{(\U_{L_1}\times\B_{L_2})}{}{-1}
 \left. +
     \frac{1}{\sqrt{l+1}} \Klm{(\U_{L_1}\times\B_{L_2})}{}{+1} \right]r\dr A_{-m}^l
   \nonumber \\ 
   &- 
   i\sqrt{\frac{2l+1}{l(l+1)}} \Klm{(\U_{L_1}\times\B_{L_2})}{}{}r^2 \left.\Delta_l\left(\frac{C_{-m}^l}{r}\right) \right\},
  \\
   \mathcal{F}_L\left(L_1,L_2,r\right) &=
     -\frac{\sqrt{4\pi}}{r^2} \partial_r \left\{ r^2 \sum_{L}
       \sum_{\nu_1 = -1}^{1} \sum_{\nu_2 = -1}^{1}
     \right. 
             \left(\mathbf{U}_{L_1}\times\mathbf{B}_{L_2}\right)^m_{l,l+\nu_1}
       B^{-m}_{l,l+\nu_2}
       \left.\mathcal{J}^{0,0,1}_{l,m,\nu_1,l,-m,\nu_2} \right\}\, .
  \label{eq:flux_full_expr}
\end{align}
The laplacian formula used for $\mathcal{D}_1$ in the vectorial
spherical harmonics basis is given in equation \eqref{eq:lapla_vect}. In the
production term we simply made use of the basis transformation
\eqref{eq:base_change}. Finally, the expressions for $\mathcal{D}_2$
and the flux term need some intermediate steps to be properly
explained. These details are
given in Appendix \ref{sec:simpl-magn-energy} for the flux of magnetic
energy, and the same procedure may be applied in the case of the second
diffusive term.

\subsection{Simplification of the magnetic energy flux}
\label{sec:simpl-magn-energy}
The flux of magnetic energy can be simplified, if one notes that it has the general form
\begin{equation*}
  F = \int_S \bnab\cdot\mathbf{X}\dint{\Omega}{}
  \hspace{1cm}\mbox{and}\hspace{1cm} \mathbf{X} = \sumlm{} \left\{ \mathcal{A}^l_m\Rlm{} + \mathcal{B}^l_m\Slm{}
  + \mathcal{C}^l_m\Tlm{}\right\} \, .
\end{equation*}
Then, one can easily deduce that
\begin{equation*}
  F = \int_S \bnab\cdot\mathbf{X}\dint{\Omega}{} =
  \frac{\sqrt{4\pi}}{r^2}\partial_r\left(r^2\mathcal{A}^0_0\right) .
\end{equation*}
Recalling from the system \eqref{eq:base_change} that $\mathbf{R}^0_0 = -
\mathbf{Y}^{0}_{0,1}$, and if one assumes that
$\mathbf{X}=\mathbf{X}_1\times\mathbf{X}_2$, one obtains, for an
integral similar to the magnetic energy flux:
\begin{equation*}
  \int_S
  \bnab\cdot\left(\mathbf{X}_1\times\mathbf{X}_2\right)\dint{\Omega}{}
  = -\frac{\sqrt{4\pi}}{r^2}\partial_r\left(r^2X^0_{0,1}\right)  =
  -\frac{\sqrt{4\pi}}{r^2}\partial_r\left\{ r^2
    \sum_{l=0}^{+\infty}\sum_{m=-l}^l\sum_{\nu_1,\nu_2}
    X_{1:l,l+\nu_1}^{m}X_{2;l,l+\nu_2}^{-m} \mathcal{J}^{0,0,1}_{l,m,\nu_1,l,-m,\nu_2} \right\}.
\end{equation*}

\subsection{On primary and secondary families}
\label{sec:prim-second-famil}

Previous studies on dynamos in stars shed light on the important
distinction of \textit{primary} (or \textit{dipolar}, antisymmetric
with respect to the equator) and
\textit{secondary} (or \textit{quadrupolar}, symmetric with respect to
the equator) families of magnetic
field. For a vector $\mathbf{X}=\sumlm{}\left\{\mathcal{A}^l_m\Rlm{} + \mathcal{B}^l_m\Slm{}
  + \mathcal{C}^l_m\Tlm{}\right\}$, \citet{Roberts:1972wu} define the primary family as
\begin{eqnarray*}
  \mathbf{X}^p &=&
  \mathcal{A}^{m+1}_{m}\mathbf{R}^{m}_{m+1} +
  \mathcal{B}^{m+1}_m\mathbf{S}^{m}_{m+1} +
  \mathcal{C}^{m}_m\mathbf{T}^{m}_{m} + 
  \mathcal{A}^{m+3}_m\mathbf{R}^{m}_{m+3} +
  \mathcal{B}^{m+3}_m\mathbf{S}^{m}_{m+3} +
  \mathcal{C}^{m+2}_m\mathbf{T}^{m}_{m+2}  
  + \dotsc
\end{eqnarray*}
and the secondary family as
\begin{eqnarray*}
  \mathbf{X}^s &=&
  \mathcal{A}^{m}_{m}\mathbf{R}^{m}_{m} +
  \mathcal{B}^{m}_m\mathbf{S}^{m}_{m} +
  \mathcal{C}^{m+1}_m\mathbf{T}^{m}_{m+1} + 
  \mathcal{A}^{m+2}_m\mathbf{R}^{m}_{m+2} +
  \mathcal{B}^{m+2}_m\mathbf{S}^{m}_{m+2} +
  \mathcal{C}^{m+3}_m\mathbf{T}^{m}_{m+3}  
  + \dotsc
\end{eqnarray*}
It can also be easily shown that
in the $\left(\mathbf{Y}^m_{l,l+\nu}\right)_{\nu = -1,0,1}$ basis, a
primary field always satisfies $l+m+\nu$ even, and a
secondary field always satisfies $l+m+\nu$ odd. 

Note that the vectorial product depends on the $3$-j Wigner 
\begin{eqnarray*}
\left(
\begin{array}{ccc}
   l_1+\nu_1 & l_2+\nu_2 & l+\nu  \\
   0 & 0 & 0
\end{array}
\right).
\end{eqnarray*}
Recalling that $m_1+m_2=m$, this $3$-j Wigner is zero if $l_1+\nu_1+l_2+\nu_2+l+\nu =
(l_1+\nu_1+m_1)+(l_2+\nu_2+m_2)+(l+\nu+m)-2m$ is odd. Consequently, in
order to have a non-zero $3$-j Wigner, if $\U$
and $\B$ are from different families, their vectorial product is a
secondary field; and if they are from the same family, their
vectorial product is a primary field. If
$\mathbf{C}=\mathbf{A}\times\mathbf{B}$, this means that
\begin{equation}
\left.
\begin{array}{ccc}
  \mathbf{A}^p & \times & \mathbf{B}^s \\ 
  \mathbf{A}^s & \times & \mathbf{B}^p
\end{array}
\right\} \rightarrow \mathbf{C}^s\;
\mbox{    and   }
\left.
\begin{array}{ccc}
  \mathbf{A}^p & \times & \mathbf{B}^p \\ 
  \mathbf{A}^s & \times & \mathbf{B}^s
\end{array}
\right\} \rightarrow \mathbf{C}^p \, ,
  \label{eq:families_rules}
\end{equation}
where the superscripts $p$ and $s$ stand for primary and secondary.
This was already acknowledged by \citet{McFadden:1991bw} and \citet{Gubbins:1993iia}.

\section{Numerical validation}
\label{sec:numerical-validation}

In order to validate the way we implemented in the ASH code the complex interactions between
spherical harmonics, we compared an analytic calculation for a
simple setup with numerical results. We summarize here those
calculations.

We start from a mixed $m=0,\pm 1$ and $l=1$ state for the magnetic field,
and an $(l,m)=(2,\pm 1)$ state for the velocity field. We initialize
the magnetic field in the following way:
\begin{eqnarray*}
  \mathbf{B} = \frac{2R_\odot^2 R_b}{r^3} \left( \mathbf{R}_1^{0} +
    \frac{1}{2}\mathbf{R}_1^{1} - \frac{1}{2}\mathbf{R}_1^{-1} \right)
  - \frac{R_\odot^2R_b}{r^3} \left( \mathbf{S}_1^{0} +
    \frac{1}{2}\mathbf{S}_1^{1} - \frac{1}{2}\mathbf{S}_1^{-1} \right)
  + \frac{\beta^2}{rR_\odot^2} \left( \mathbf{T}_1^{0} +
    \frac{1}{2}\mathbf{T}_1^{1} - \frac{1}{2}\mathbf{T}_1^{-1} \right),
\end{eqnarray*}
where $R_\odot$ is the solar radius, $R_b$ is our inner boundary
radius, $R_t$ is our outer boundary radius, and $\beta=(R_t-r)^2(r-R_b)^2$.
The velocity is initialized by:
\begin{eqnarray*}
  \bar{\rho}\mathbf{U} = \frac{3\beta^2}{r^2R_\odot^4}\left(
    \mathbf{R}_2^{1} - \mathbf{R}_2^{-1} \right) + \frac{3\chi
    r}{2R_\odot^2} \left(
    \mathbf{S}_2^{1} - \mathbf{S}_2^{-1} \right) + \frac{\beta^2}{2rR_\odot^2}\left(
    \mathbf{T}_2^{1} - \mathbf{T}_2^{-1} \right),
\end{eqnarray*}
where $\chi =
(R_t-r)^2(r-R_b)^2(R_t+R_b-2r)/(r^2R_\odot^2)$. Rewriting those fields
in the conventional spherical harmonics writing, we may
calculate the axisymmetric components of the vectorial product
$\bar{\rho}\mathbf{U}\times\mathbf{B}$ and obtain
\begin{eqnarray}
\bar{\rho}\mathbf{U}\times\mathbf{B} = \left(
  \begin{array}{c}
    -\frac{2\gamma_r}{5}\sqrt{\frac{4\pi}{7}}Y_3^0 -
    \frac{3\gamma_r}{5}\sqrt{\frac{4\pi}{3}}Y_1^0\\
    -\frac{2\gamma_\theta^2}{15}\sqrt{\frac{4\pi}{7}}\partial_\theta Y_3^0 -
    \left( \frac{\gamma_\theta^2}{5} +
      \gamma_\theta^1\right)\sqrt{\frac{4\pi}{3}} \partial_\theta Y _1^0 \\
     -\frac{2\gamma_\varphi^2}{15}\sqrt{\frac{4\pi}{7}}\partial_\theta Y_3^0 -
    \left( \frac{\gamma_\varphi^2}{5} +
      \gamma_\varphi^1\right)\sqrt{\frac{4\pi}{3}} \partial_\theta Y _1^0 
  \end{array}
\right),  
\label{eq:vect_prod_analytic}
\end{eqnarray}
where the $\gamma$ coefficients are defined by:
\begin{eqnarray*}
  \gamma_r &=& \frac{9\sqrt{5}\chi\beta^2}{8\pi R_\odot^2} +
  \gamma_\theta^1, \\
  \gamma_\theta^1 =\frac{3\sqrt{5}\beta^2R_b}{8\pi r^4} \mbox{
  }&,&\mbox{   } \gamma_\theta^2 = \frac{9\sqrt{5}\beta^5}{8\pi r^3
    R_\odot^6} - \gamma_\theta^1 \\
  \gamma_\varphi^1 =\frac{9\sqrt{5}\chi R_b}{8\pi r^2} \mbox{
  }&\mbox{and}&\mbox{   } \gamma_\varphi^2 = -\frac{9\sqrt{5}\beta^3 R_b}{8\pi r^5
    R_\odot^2} - \gamma_\varphi^1 .
\end{eqnarray*}
These coefficients match exactly the outputs from the code \referee{(table \ref{tab:tab1})}. The
production and flux terms in equation \eqref{eq:ME_terms} are then
simple scalar products involving the vectorial product (\ref{eq:vect_prod_analytic}). They
also have been checked by comparison with the analytical
calculation.

\begin{deluxetable*}{cccc}
  \tablecaption{Analytical and numerical values of the vectorial product
    $\mathbf{U}\times\mathbf{B}$ (validation case, see Sect. \ref{sec:valid-meth-simple}).\label{tab:tab1}}
  \tablecomments{The values are evaluated at $r=0.84\, R_\odot$. The
    expressions for the $\gamma$ coefficients are given in Appendix
    \ref{sec:numerical-validation}. Numerical results are given with
    $15$ significant digits, \textit{i.e.} up to the numerical accuracy.}
  \tablecolumns{4}
  \tabletypesize{\scriptsize}
  \tablehead{
    \colhead{SH Mode} &
    \colhead{Analytical Expression} &
    \colhead{Analytical Value} &
    \colhead{Code Output}
  }
  \startdata 
  $(1,0)$ &
  $   \left(
    \begin{array}{c}
      -\frac{3\gamma_r}{5}\sqrt{\frac{4\pi}{3}}\\
      -\left( \frac{\gamma_\theta^2}{5} +
        \gamma_\theta^1\right)\sqrt{\frac{4\pi}{3}} \\
      \left( \frac{\gamma_\varphi^2}{5} +
        \gamma_\varphi^1\right)\sqrt{\frac{4\pi}{3}}
    \end{array}
  \right)$
  &
  $   \left(
    \begin{array}{c}
      -12949906405.7406 \\
      -12949960422.6225 \\
      -0.00337401832383
    \end{array}
  \right)$
  &
  $   \left(
    \begin{array}{c}
      -12949906405.7402 \\
      -12949960422.6225 \\
      -0.00337401832383
    \end{array}
  \right)$
  \\
  $(2,0)$ &
  $   \left(
    \begin{array}{c}
      0 \\
      0 \\
      0
    \end{array}
  \right)$
  &
  $   \left(
    \begin{array}{c}
      0.0 \\
      0.0 \\
      0.0
    \end{array}
  \right)$
  &
  $   \left(
    \begin{array}{c}
      0.0 \\
      0.0 \\
      0.0
    \end{array}
  \right)$
  \\
  $(3,0)$ &
  $   \left(
    \begin{array}{c}
      -\frac{2\gamma_r}{5}\sqrt{\frac{4\pi}{7}} \\
      -\frac{2\gamma_\theta^2}{15}\sqrt{\frac{4\pi}{7}} \\
      \frac{2\gamma_\varphi^2}{15}\sqrt{\frac{4\pi}{7}} 
    \end{array}
  \right)$
  & 
  $   \left(
    \begin{array}{c}
      -5651802509.22854 \\
      \phm{-}  3767844764.58569 \\
      -0.00147254232049
    \end{array}
  \right)$
  &
  $   \left(
    \begin{array}{c}
      -5651802509.22835 \\
      \phm{-}  3767844764.58568 \\
      -0.00147254232049
    \end{array}
  \right)$
  \enddata
\end{deluxetable*}

\end{document}